\documentclass{nature}
\usepackage{amsfonts}
\usepackage{amssymb}
\usepackage{amsmath}
\usepackage{bm}
\usepackage{braket}
\usepackage{xcolor}
\usepackage{graphicx}
\usepackage{textcomp,gensymb}

\makeatletter
\let\saved@includegraphics\includegraphics
\AtBeginDocument{\let\includegraphics\saved@includegraphics}
\renewenvironment*{figure}{\@float{figure}}{\end@float}
\makeatother
\usepackage[hidelinks]{hyperref}
\usepackage[T1]{fontenc}

\graphicspath{{./figures/}}
\newcommand{\parallelsum}{\mathbin{\!/\mkern-5mu/\!}}
\DeclareMathAlphabet\mathbfcal{OMS}{cmsy}{b}{n}

\newcommand{\changemyfigure}{\linespread{1.3}\selectfont{}}

\newcommand{\onlinecite}[1]{\hspace{-1 ex} \nocite{#1}\citenum{#1}}

\begin{document}

\title{Exploring Superconductivity under Strong Coupling with the Vacuum Electromagnetic Field}

\author{A.~Thomas$^1$, E.~Devaux$^1$, K.~Nagarajan$^1$, T.~Chervy$^1$, M.~Seidel$^1$, D.~Hagenm\"uller$^1$, S.~Sch\"utz$^1$, J.~Schachenmayer$^{1}$, C.~Genet$^1$, G.~Pupillo$^{1*}$ \& T.~W. Ebbesen$^{1*}$}

\maketitle

\vspace{-5mm}

\begin{affiliations}
 \item ISIS (UMR 7006) \& icFRC, University of Strasbourg and CNRS, 67000 Strasbourg, France
\end{affiliations}

\begin{abstract}
Light-matter interactions have generated considerable interest as a means to manipulate material properties~\cite{Laussy2010,Cotlet2016,Fausti2011,Hutchison2012,Orgiu2015,Feist2015,Schachenmayer2015,Mitrano2016,Ebbesen2016,Ebbesen2016patent,Thomas2016,Zhong2017,Hagenmuller2017,Sentef2018,Nagarajan2018,Stranius2018,Paravicini2019,Cortese2019,Thomas2019,KenaCohen2019,Hagenmuller2019,Schlawin2019,Curtis2019}. Light-induced superconductivity has been  demonstrated~\cite{Fausti2011,Mitrano2016} using pulsed lasers. An attractive alternative possibility is to exploit strong light-matter interactions arising by coupling phonons to the vacuum electromagnetic field of a cavity mode as has been suggested~\cite{Ebbesen2016,Ebbesen2016patent} and theoretically studied~\cite{Sentef2018,Hagenmuller2019}. Here we explore this possibility for two very different superconductors, namely YBCO (YBa$_2$Cu$_3$O$_{6+\delta}$)~\cite{Wu1987} and Rb$_3$C$_{60}$~\cite{Tanigaki1991,Gunnarsson2005}, coupled to surface plasmon polaritons, using a novel cooperative effect based on the presence of a strongly coupled vibrational environment allowing efficient dressing of the otherwise weakly coupled phonon bands of these compounds~\cite{Lather2019}. By placing the superconductor-surface plasmon system in a SQUID magnetometer, we find that the superconducting transition temperatures {\bf($T_c$}) for both compounds are modified in the absence of any external laser field. For YBCO, {\bf $T_c$} decreases from 92 K to 86 K while for Rb$_3$C$_{60}$, it increases from 30 K to 45 K at normal pressures. In the latter case, a simple theoretical framework is provided to understand these results based on an enhancement of the electron-phonon coupling. This proof-of-principle study opens a new tool box to not only modify superconducting materials but also to understand the mechanistic details of different superconductors.
\end{abstract}

%\section{Introduction} 

Light-matter strong coupling is typically achieved by placing a material in the confined field of an optical mode such as a surface plasmon polariton (SPP) resonance. SPPs are evanescent waves propagating along a metal-dielectric interface, which allow tight confinement of the electromagnetic field. Strong coupling leads to the formation of hybrid light-matter polaritonic states separated by what is known as the Rabi splitting. Even in the dark, this Rabi splitting has a finite value which is due to the interaction with the vacuum electromagnetic field, and it can be collectively enhanced with the number of coupled oscillators in the material. 
 
Strong coupling has been shown to modify material properties and lead to an enhancement of charge~\cite{Orgiu2015,Hagenmuller2017,Nagarajan2018,Paravicini2019} and energy~\cite{Schachenmayer2015,Feist2015,Zhong2017} transport. Furthermore, the strong coupling of molecular vibrations shows remarkable changes in chemical reactivity landscapes~\cite{Thomas2016,Thomas2019,KenaCohen2019} and it is therefore natural to investigate whether strong coupling of crystal phonons can lead to modifications of superconductivity~\cite{Ebbesen2016,Hagenmuller2019}.  

\begin{figure}
\centerline{\includegraphics[width=0.5\columnwidth]{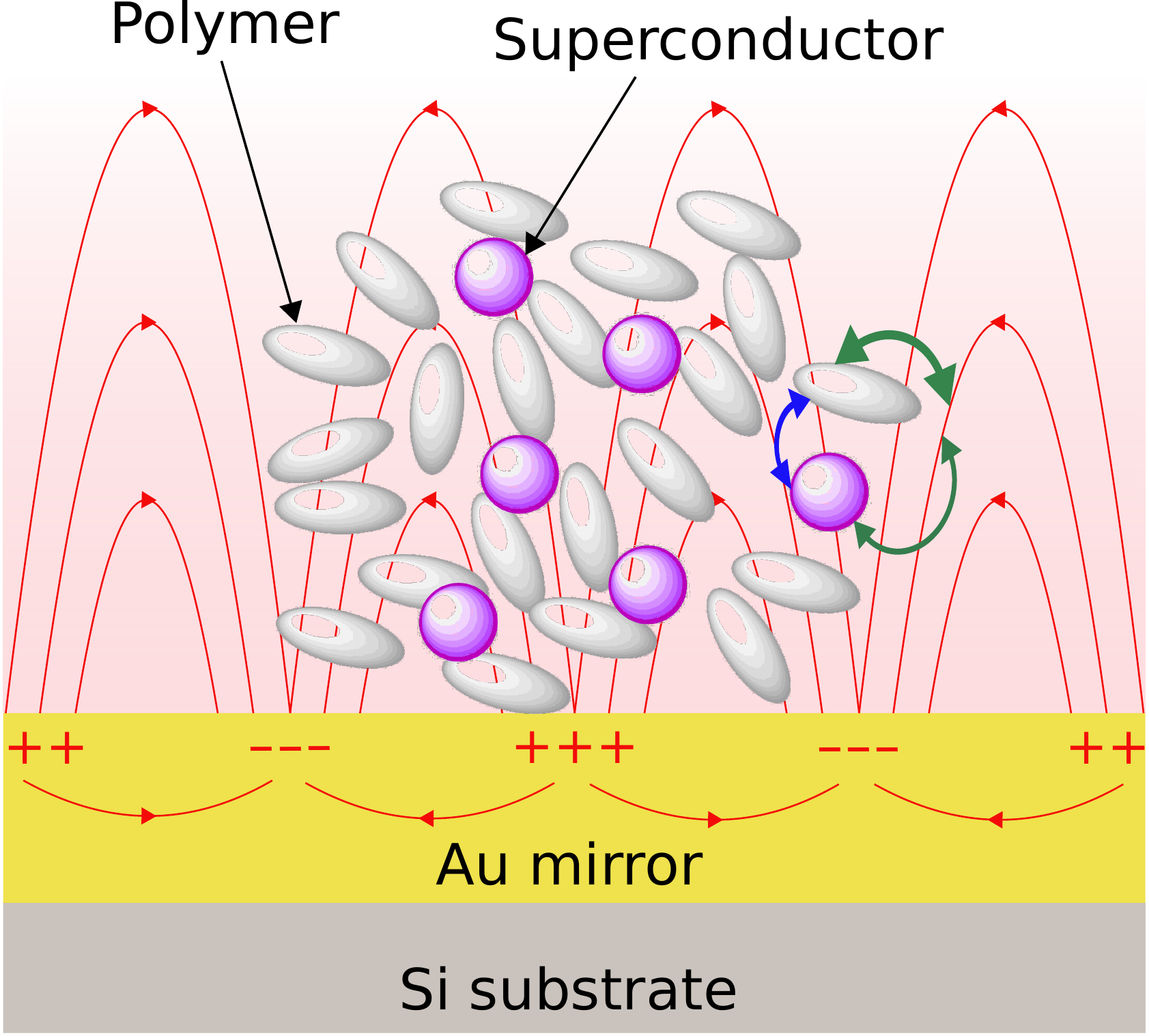}}
\caption{\changemyfigure{}{\bf Illustration of the cooperative strong coupling mechanism in the present study.} Superconductors (purple spheres) are embedded in a dielectric polymer matrix (grey ellipses) and deposited on a thin metal layer (Au). The metal-dielectric interface generates surface plasmon polaritons with electric field represented by the red lines. While phonons in the superconductor are only weakly coupled to the field of the surface plasmon polaritons (thin green arrow), the latter is strongly coupled to the polymer vibrations (thick green arrow). Superconductor phonons and polymer vibrations also directly interact via dipole-dipole coupling (blue arrow).}  
\label{fig_sketch}
\end{figure}

For the purpose of this study, we chose two well-known type II superconductors (SCs), the unconventional YBCO and  Rb$_3$C$_{60}$~\cite{Wu1987,Tanigaki1991,Gunnarsson2005}. The challenge in testing such compounds comes from the weak transition dipole moment of the phonons involved in superconductivity. As we have shown recently, it is possible to exploit a vibrational environment consisting of auxiliary oscillators~\cite{Lather2019,Schutz2019} that are strongly coupled to the vacuum field of a cavity, and quasi-resonant with the target oscillators that alone would not strongly couple due to a small oscillator strength. Here, we show that a similar cooperative effect (see Fig.~\ref{fig_sketch}) can lead to modifications of the $T_c$ of the two SCs, which for Rb$_3$C$_{60}$ is qualitatively captured by a modification of the electron-phonon coupling within the SC via the formation of a ``dressed'' phonon with effectively reduced frequency.

%\section{Experimental results} 

\begin{figure}
\centerline{\includegraphics[width=\columnwidth]{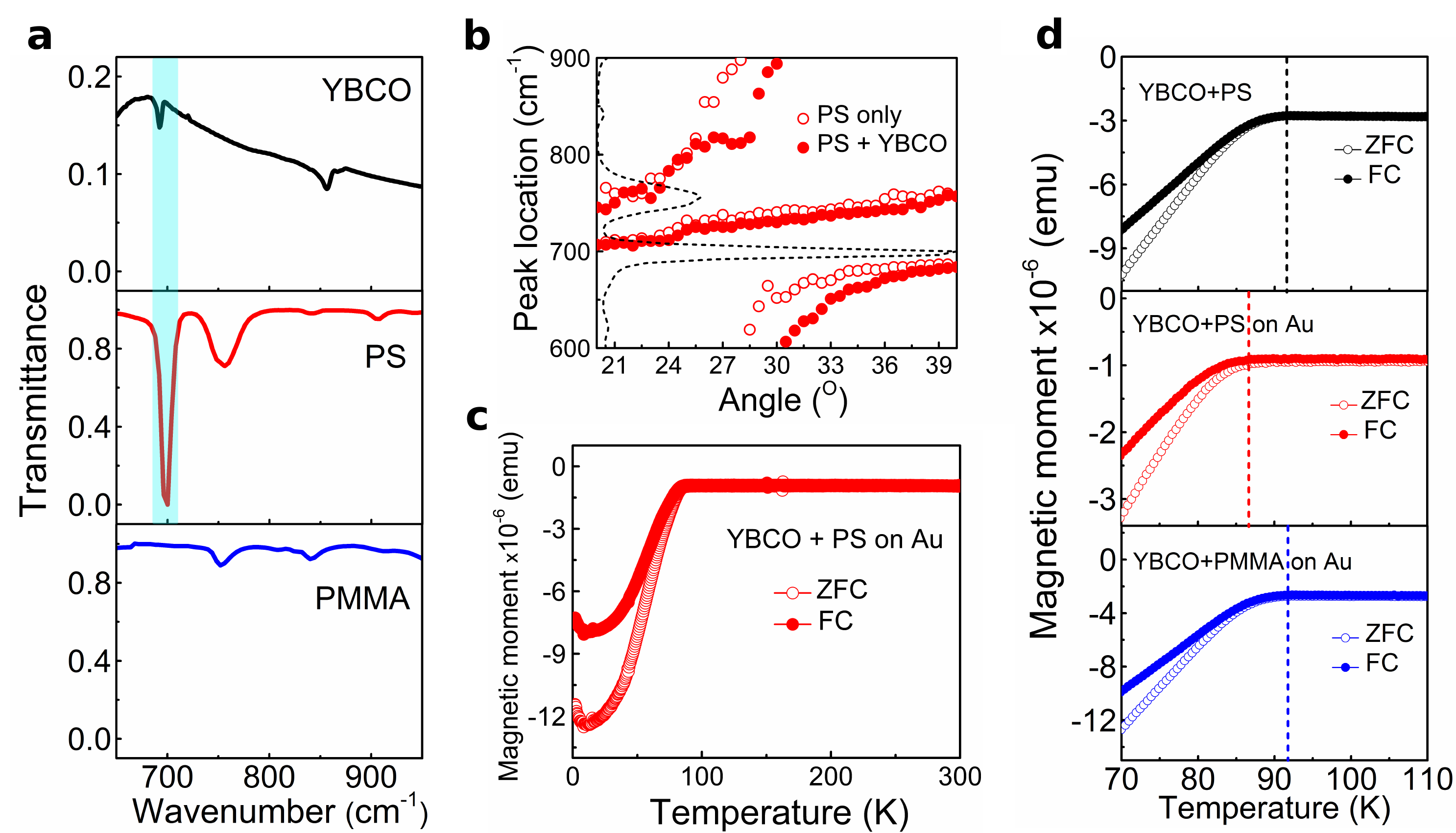}}
\caption{\changemyfigure{}{\bf Strong coupling-induced changes of $\bm{T_c}$ for YBCO.}  {\bf a}, Fourier-transform infrared (FTIR) spectrum of YBCO (top), PS (middle) and PMMA (bottom). The overlapping of the YBCO phonon mode with the PS vibrational mode is highlighted in cyan. {\bf b}, Dispersion curves showing the cooperative strong coupling of the surface plasmon mode with the PS vibration at 697 ${\rm cm}^{-1}$ which overlaps with the YBCO phonon mode (YBCO alone does not lead to strong coupling). The dashed curve represents the PS transmission spectrum, the empty circles the signal for PS alone, and the solid circles the PS+YBCO mixture. The small differences between the two curves could be due to the background dielectric contribution of YBCO. {\bf c}, Temperature-dependent magnetization of a strongly coupled YBCO film (YBCO+PS on Au) in the zero-field-cooled (ZFC) and 100 Oe field-cooled (FC) modes. {\bf d}, Comparison of the temperature-dependent magnetization for both ZFC and 100 Oe FC between 70 K and 110 K of a bare film of YBCO+PS on Si (top), strongly coupled YBCO+PS on Au (middle), and cooperatively off-resonant YBCO+PMMA on Au (bottom). While both the bare film and the cooperatively off-resonant YBCO feature a $T_c$ of 92 K, the latter is shifted to 86 K for the strongly coupled YBCO as shown by the dashed lines in the respective panels. The offsets in the $y$-axis of the magnetization curves are attributed to the substrates.}
\label{fig_exp_YBCO}
\end{figure}

The IR spectrum of YBCO powder is shown in Fig.~\ref{fig_exp_YBCO}\textbf{a} displaying two phonon peaks at the 692 cm$^{-1}$ and 856 cm$^{-1}$ that are too weak to strongly couple to a Fabry-Perot cavity or a plasmonic mode. To induce a cooperative strong coupling effect, the YBCO powder was dispersed in various polymers such as polystyrene (PS) and polymethylmethacrylate (PMMA), whose vibrational spectra are also shown in Fig.~\ref{fig_exp_YBCO}\textbf{a}, as well as polyvinyl acetate (PVAc). As can be seen, only PS has a strong peak at 697 cm$^{-1}$ that overlaps with the infrared-active phonon modes of YBCO, which are believed to be involved in superconductivity~\cite{Arai1992,Calvani1995,Bernhard2000,Liu2019}. The YBCO powder dispersed in the different polymers could have been placed inside resonant Fabry-Perot cavities but the cavity path length would have changed at low temperature, making it very hard to produce tuned samples. For this reason, we chose to spin-coat the YBCO + polymer mix onto an Au film (20 nm thick) to couple to the surface plasmons of the metal. The strong coupling can be measured by attenuated total reflection (ATR) using an FTIR (Fourier-transform infrared spectrophotometer). Figure \ref{fig_exp_YBCO}\textbf{b} gives the dispersion curves measured in this way, displaying anti-crossings typical of strong coupling. The double splitting is due to the presence of two closely lying vibrational normal modes in PS. The Rabi splitting for the coupled system at 697 cm$^{-1}$ is 60 cm$^{-1}$ which is larger than the full width half-maximum (FWHM) (16 cm$^{-1}$) of the vibrational mode and the corresponding SPP resonance, a necessary condition for the strong coupling regime. The dispersion curves are easily simulated and show that it is the PS modes that define the strong coupling (see Supplementary Information). 

The samples were then placed in gelatin capsules and introduced into a SQUID magnetometer (either a MPMS or a MPMS3, ``Quantum Design Europe''). SQUID magnetometry is a well-established technique to characterize the Meissner effect, which is one of the two hallmarks of superconductivity together with a zero resistance. The Meissner effect is demonstrated by measuring both the field-cooled (FC) and zero-field-cooled (ZFC) magnetization curves. Here the latter are typically recorded between room-temperature and 4 K at 100 Oe and are shown in Fig.~\ref{fig_exp_YBCO}\textbf{c} \& \textbf{d}. The Meissner effect is clearly visible in Fig.~\ref{fig_exp_YBCO}\textbf{c}. The bare YBCO in a PS film shows an onset of superconductivity at 92 K as expected from the literature~\cite{Wu1987} (Fig.~\ref{fig_exp_YBCO}\textbf{d}). When coated on an Au substrate, the $T_c$ decreases from 92 K to 86 K.  When the YBCO powder is dispersed in the other polymers (PMMA and PVAc), no change in $T_c$ is observed since their vibrational bands do not overlap with YBCO, as expected. The role of the metal was also checked by repeating the experiments on Pt and Ag to ensure that there was no specific role of Au in the experiments (see ``Extended Data''). The $T_c$ was again found to be shifted to 86 K on these metal films as expected since they both support plasmons in the infrared (see Supplementary Information). We find that optimal loading of YBCO in the polymer is about 10 wt$\%$, while no change in $T_c$ is observed at high loading. 

\begin{figure}
\centerline{\includegraphics[width=\columnwidth]{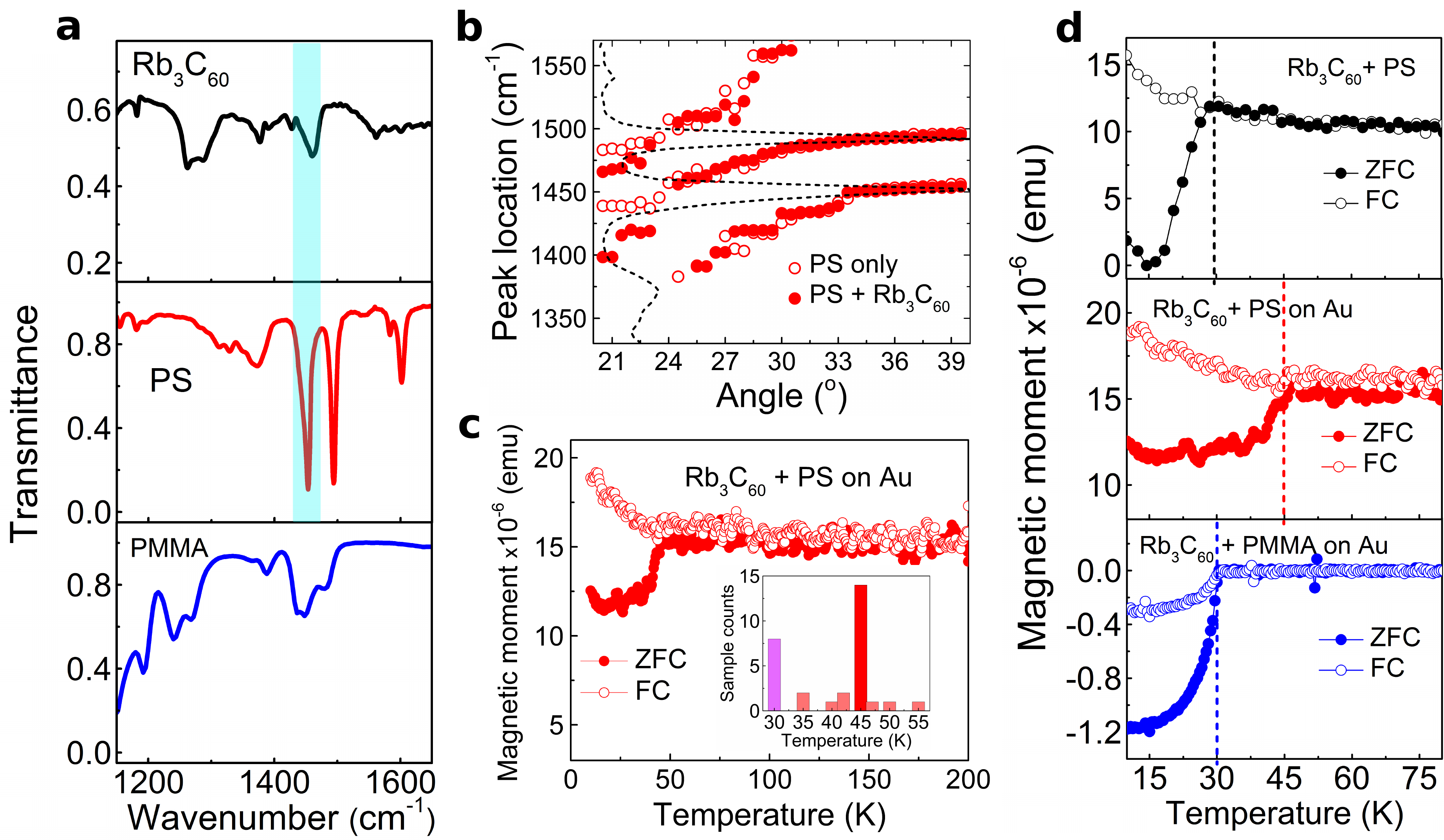}}
\caption{\changemyfigure{}{\bf Strong coupling-induced changes of $\bm{T_c}$ for Rb$\bm{_{3}}$C$\bm{_{60}}$.} {\bf a}, FTIR spectrum of ${\rm Rb_{3}C_{60}}$ powder (top), PS (middle) and PMMA (bottom). The overlapping of the ${\rm Rb_{3}C_{60}}$ phonon mode with the PS vibrational mode is highlighted in cyan. {\bf b}, Dispersion curves showing the cooperative strong coupling of the surface plasmon with the PS vibration at 1452 ${\rm cm}^{-1}$ which overlaps with the ${\rm Rb_{3}C_{60}}$ phonon mode (${\rm Rb_{3}C_{60}}$ alone does not lead to strong coupling). The dashed curve represents the PS transmission spectrum, the empty circles the signal for PS alone, and the solid circles the PS+${\rm Rb_{3}C_{60}}$ mixture. {\bf c}, Temperature-dependent magnetization of a strongly coupled ${\rm Rb_{3}C_{60}}$ film (${\rm Rb_{3}C_{60}}$+PS on Au) for ZFC and 100 Oe FC between of 10 K and 200 K. The inset shows the statistics of the observed $T_c$ for the various samples studied. {\bf d}, Comparison of the temperature-dependent magnetization for ZFC and 100 Oe FC between 10 K and 80 K of the bare film of ${\rm Rb_{3}C_{60}}$+PS on Si (top), strongly coupled ${\rm Rb_{3}C_{60}}$+PS on Au (middle), and cooperatively off-resonant ${\rm Rb_{3}C_{60}}$+PMMA on Au (bottom). For both the bare film and cooperatively off-resonant ${\rm Rb_{3}C_{60}}$ the $T_c$ is at 30 K, while it is increased to 45 K for the strongly coupled ${\rm Rb_{3}C_{60}}$ as shown by the dashed lines in the respective panels. The magnetization offsets are attributed to the substrates and the Pyrex tube.}  
\label{fig_exp_Rb}
\end{figure}

Next Rb$_3$C$_{60}$ was investigated using the same approach. Fullerene SCs are extremely air and humidity sensitive due to the radical nature of the reduced C$_{60}$, so special precautions had to be taken in the preparation of the samples. As described in the ``Methods'' section, the SC was prepared by mixing C$_{60}$ and Rb in a glovebox under inert atmosphere and baked in a sealed ampoule at 400 \degree C. The IR spectrum of Rb$_3$C$_{60}$ is shown in Fig.~\ref{fig_exp_Rb}\textbf{a} with a distinct but weak peak at 1460 cm$^{-1}$. This is one of the on-ball modes of C$_{60}$ (Hg symmetry) that are thought to play an important role in the superconducting properties of Rb$_3$C$_{60}$~\cite{Gunnarsson2005}. PS also features vibrational bands that match this mode (see Fig.~\ref{fig_exp_Rb}\textbf{a}). Hence the Rb$_3$C$_{60}$ powder was dispersed in a PS solution and spin-coated on a 20 nm Au film in the glovebox in order to see if the cooperative coupling modifies the $T_c$. The dispersion curves in Fig.~\ref{fig_exp_Rb}\textbf{b} measured by ATR show a double splitting due to two close lying PS modes. The mode overlapping with the Rb$_3$C$_{60}$ peak at 1460 cm$^{-1}$ yields a Rabi splitting of 48 cm$^{-1}$, much larger than the FWHM of the vibrational band (10 cm$^{-1}$). The samples were placed in sealed Pyrex tubes, introduced into the SQUID magnetometer and the FC and ZFC curves were recorded. As can be seen in Fig.~\ref{fig_exp_Rb}\textbf{c}, the strongly coupled polymer plus Rb$_3$C$_{60}$ film on Au shows a $T_c$ of 45 K. This is an increase of 50\% over the normal $T_c$ of 30 K shown in Fig.~\ref{fig_exp_Rb}\textbf{d}. As in the case of YBCO, PMMA had no effect on $T_c$ because it has no vibrational bands that overlap with the SC. It should be noted that the $T_c$ of the strongly coupled system varies slightly from sample to sample. Nevertheless, it strongly peaks at 45 K with some outlyers as shown in the inset of Fig.~\ref{fig_exp_Rb}\textbf{c}. Out of the 33 samples tested 10 remained at 30 K which we attribute to delamination of the polymer film from the Au substrate during cooling, as this could be seen even with the eye when the samples were removed from the SQUID. The optimal loading of Rb$_3$C$_{60}$ in the polymer was found to be around 6 wt$\%$, a point that is discussed in the theoretical part below. We note that the existence of a single drop in the magnetization curves of Figs.~\ref{fig_exp_YBCO} and \ref{fig_exp_Rb} indicates that the entire superconducting volume fraction in the materials features a modified $T_c$ in the presence of cooperative strong coupling.  

%\section{Theoretical model} 

%We now introduce a theoretical model providing a qualitative explanation for an enhancement of the electron-phonon coupling in the presence of cooperative strong coupling between SC phonons and SPPs. The electron-phonon coupling governs the scattering of electrons across the Fermi surface giving rise to Cooper pairing~\cite{Bardeen1957} and conventional superconductivity in, e.g., Rb$_3$C$_{60}$~\cite{Gunnarsson2005}.  The associated $T_{c}$ increases with the strength of the electron-phonon coupling, which is usually described by the dimensionless parameter$\lambda$~\cite{McMillan1968,Allen1975}. 

We now introduce a theoretical model providing a qualitative explanation for the observed enhancement of $T_c$ in Rb$_3$C$_{60}$ due to cooperative strong coupling between SC phonons and SPPs. In conventional SCs such as Rb$_3$C$_{60}$~\cite{Gunnarsson2005}, the $T_{c}$ increases with the strength of the electron-phonon interaction which governs the scattering of electrons across the Fermi surface giving rise to Cooper pairing~\cite{Bardeen1957}. These processes are usually quantified by the dimensionless parameter $\lambda$~\cite{McMillan1968,Allen1975}. 

For a spherical Fermi surface and an optical phonon mode with frequency $\omega_{q}$, the parameter $\lambda\propto \int_{0}^{2k_{\rm F}} \!\! dq \, q \; \frac{V^{2}_{q}}{\omega_{q}}$ where $q$ is the phonon wave vector, $k_{\rm F}$ the Fermi wave vector, and $V_{q}$ the electron-phonon coupling matrix element. For large light-matter couplings, one could in principle exploit the formation of low frequency polaritonic states to effectively reduce $\omega_q$, which would lead to an enhancement of $\lambda$. However, sizable shifts of $\omega_q$ are usually hampered by the small coupling strength of SC phonons to the (vacuum) electromagnetic field~\cite{Sentef2018,Hagenmuller2019}. Furthermore, while the largest shifts occur close to the resonance at optical wave-vectors $q\sim \omega_{q}/c \ll k_{\rm F}$ ($c$ is the speed of light), $V_q$ is in general a slowly varying function of $q$~\cite{Giustino2017}, which implies that large enhancements of $\lambda$ can be mainly engineered by a reduction of $\omega_q$ at large wave-vectors $q\sim k_{\rm F}$.  

Here, we propose a mechanism to induce large phonon shifts at wave vectors $q\sim k_{\rm F}$ relevant for superconductivity via the auxiliary polymer. When the polymer and the SC phonons are quasi-resonant, we find that a strong coupling of the polymer to the far off-resonant electromagnetic field for which $q\sim k_{\rm F} \gg \omega_{q}/c$ leads to a low-frequency polariton containing a finite SC phonon weight due to direct dipole-dipole interactions of the latter with the polymer. This results in an enhancement of $\lambda$ that increases with the polymer coupling strength to SPPs. 

\begin{figure}[htp]
\centerline{\includegraphics[width=0.92\columnwidth]{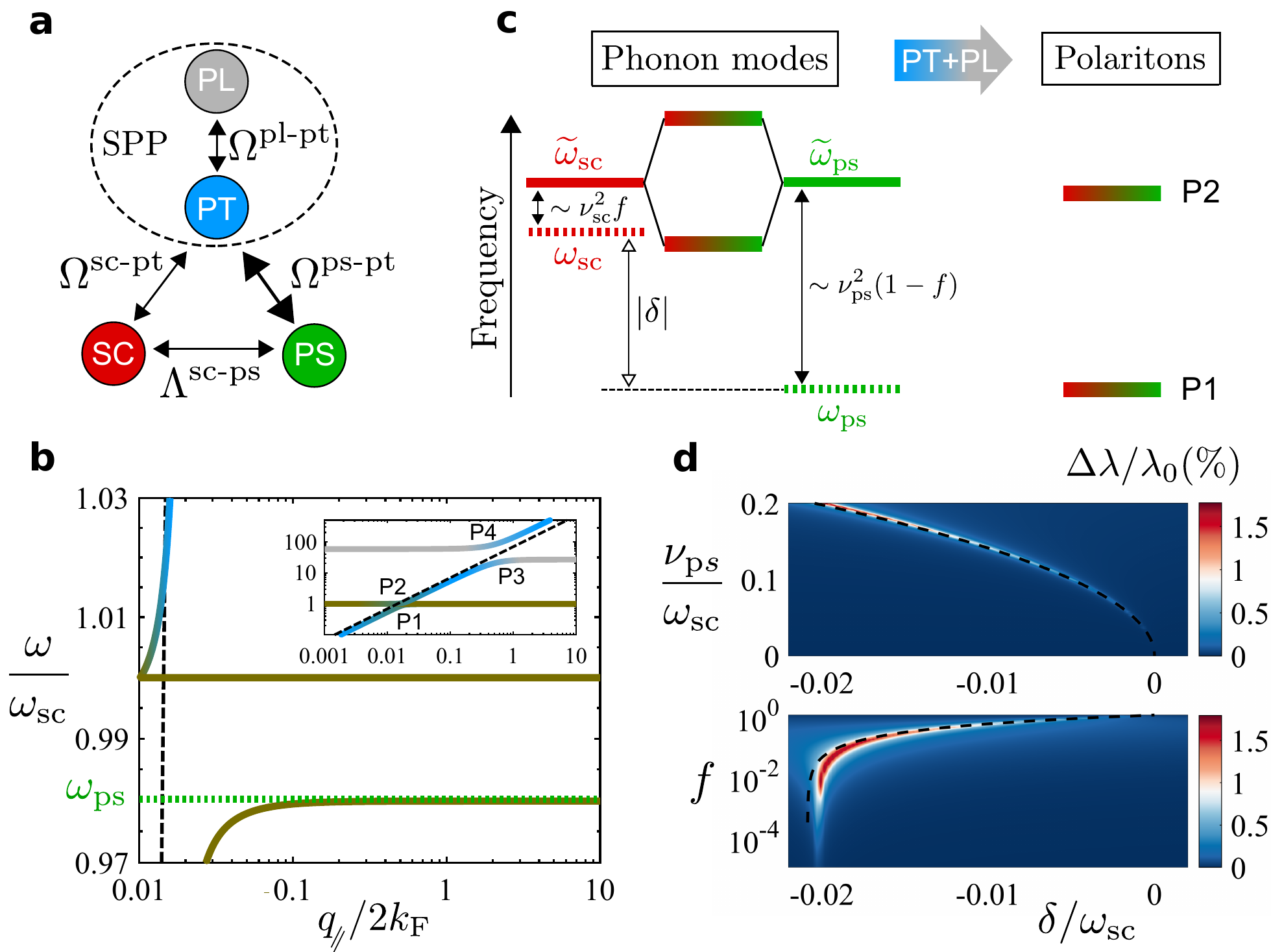}}
\caption{\changemyfigure{}{\bf Theoretical model.} \textbf{a}, Sketch of the four relevant bosonic modes of the model and their coupling strengths: SC (red) and PS (green) phonons are coupled among each other and to photons (PT), which are hybridized with plasmons (PL) to SPPs (dashed circle). \textbf{b}, Polariton energy dispersion ($f=0.02$, $\nu_{\rm ps}=0.2 \omega_{\rm sc}$, $\delta\approx -0.02\omega_{\rm sc}$) with mode admixtures (line colors correspond to {\bf a}). The brownish color of P1 indicates a mixture of SC and PS with a SC phonon weight $\approx 49.6\%$ for $q_{\parallelsum} =2k_{\rm F}$. The photon dispersion $\omega = q_{\parallelsum} c/\sqrt{\epsilon}$ is depicted as a black dashed line. \textbf{c}, Sketch of the energy levels for $q_{\parallelsum} \sim k_{\rm F}$, including energy shifts of phonons due to dipolar self-interactions ($\omega_{{\rm sc}/{\rm ps}} \to \widetilde{\omega}_{{\rm sc}/{\rm ps}}$). Enhancement of $\lambda$ can be achieved for a P1 with frequency smaller than $\omega_{\rm sc}$ ($\delta\equiv\omega_{\rm ps}-\omega_{\rm sc} < 0$) and substantial SC weight ensured by $\widetilde{\omega}_{\rm sc}=\widetilde{\omega}_{\rm ps}$. This can occur when the collective dipole moment of PS phonons $\sim \nu_{\rm ps}\sqrt{1-f}$ exceeds that of SC phonons $\sim \nu_{\rm sc}\sqrt{f}$. \textbf{d}, $\Delta\lambda/\lambda_{0}$ (\%) versus $\delta$, $\nu_{\rm ps}$, and $f$, for $f=0.02$ (top) and $\nu_{\rm ps}=0.2\omega_{\rm sc}$ (bottom). The dashed lines corresponds to $\widetilde{\omega}_{\rm sc}=\widetilde{\omega}_{\rm ps}$.}  
\label{fig_theory}
\end{figure}

We consider a film described as an effective homogeneous dielectric medium with background dielectric constant of the polymer $\epsilon$, and two optical phonons with bare frequencies $\omega_{\rm sc}$ and $\omega_{\rm ps}$, associated to an on-ball mode of C$_{60}$ and the quasi-resonant mode of PS, respectively. Dipole-dipole self-interactions for each type of phonons provide a depolarization shift of the bare frequencies: $\widetilde{\omega}_{\rm sc}=\sqrt{\omega^{2}_{\rm sc}+\nu^{2}_{\rm sc}f}$ and $\widetilde{\omega}_{\rm ps}=\sqrt{\omega^{2}_{\rm ps}+\nu^{2}_{\rm ps}(1-f)}$, where $f$ is the SC volume fraction in the film (loading) and $\nu_{\rm sc,ps}$ are associated to the phonon oscillator strengths. These shifted frequencies correspond to the minima of a transmission spectrum measured at normal incidence (top of the ``polaritonic gaps''~\cite{Anappara2009,George2016}). The two types of phonons further interact with each other via direct dipole-dipole coupling with strength $\Lambda^{\rm sc-ps} \propto \nu_{\rm sc}\nu_{\rm ps}\sqrt{f(1-f)}$. Photons are coupled to SC and PS phonons with strengths $\Omega^{{\rm sc}- {\rm pt}}_{q_{\parallelsum}} \propto \nu_{\rm sc} \sqrt{f}$ and $\Omega^{{\rm ps}- {\rm pt}}_{q_{\parallelsum}} \propto \nu_{\rm ps} \sqrt{(1-f)}$, respectively, and to plasmons in the metal (plasma frequency $\omega_{\rm pl}$) with strength $\Omega^{\rm pl-pt}_{q_{\parallelsum}} \propto \sqrt{\omega_{\rm pl}}$ (Fig.~\ref{fig_theory}\textbf{a}). All the modes coupled to light are quasi-2d ``bright'' modes with a given in-plane wave vector $q_{\parallelsum}$ (parallel to the metal-dielectric interface), superpositions of the 3d modes (phonons or plasmons) with different wave vectors in the out-of-plane direction. The resulting Hamiltonian comprises four coupled harmonic oscillators (see ``Methods''), whose eigenvalues provide the polariton frequencies shown in Fig.~\ref{fig_theory}\textbf{b}. At large wave vectors $q_{\parallelsum}\sim 2 k_{\rm F}$, the spectrum consists of a light-like branch (P4) with $\omega \sim c q_{\parallelsum}\sqrt{\frac{1+\epsilon}{\epsilon}}$ ($c$ is the speed of light in vacuum), a surface plasmon branch (P3) with $\omega \sim \frac{\omega_{{\rm pl}}}{\sqrt{1+\epsilon}}$, as well as two ``dressed'' phonons (P1 and P2) at much lower frequencies.

We find that the electron-phonon coupling parameter $\lambda$ is enhanced by the formation of the lowest polariton P1. Motivated by the dispersion curves (see Supplementary Information), we assume $\nu_{\rm ps} > \nu_{\rm sc}$. For small $f$ the collective dipole moment of PS phonons $\sim \nu_{\rm ps}\sqrt{1-f}$ thus largely exceeds that of the SC phonons $\sim\nu_{\rm sc}\sqrt{f}$, and the light-matter coupling is dominated by the polymer. In this case, the frequency of P1 at large $q_{\parallelsum}\sim 2 k_{\rm F}$ is close to the bottom of the PS polaritonic gap $\omega_{\rm ps}$, just as in polar crystals where the lower phonon polariton approaches the transverse optical phonon frequency at large wave vectors. Furthermore, when the two shifted phonons are in resonance ($\widetilde{\omega}_{\rm sc}=\widetilde{\omega}_{\rm ps}$), the PS phonon frequency $\omega_{\rm ps}$ lies below $\omega_{\rm sc}$, i.e., $\delta\equiv \omega_{\rm ps}- \omega_{\rm sc} < 0$ (see Fig.~\ref{fig_theory}\textbf{c}), and one obtains an effective redshift of SC phonons at large wave vectors if P1 contains a significant SC phonon weight. The latter is achieved via the hybridization of SC and PS phonons due to direct dipole-dipole interactions, which is maximum ($50\%-50\%$) when $\widetilde{\omega}_{\rm sc}=\widetilde{\omega}_{\rm ps}$. We expect that this latter resonance condition is relaxed in the experiments due to the finite phonon linewidth that is not included in the model. Figure~\ref{fig_theory}\textbf{d} shows the relative change $\frac{\Delta \lambda}{\lambda_{0}}\equiv \frac{(\lambda-\lambda_{0})}{\lambda_{0}}$ as a function of the detuning $\delta$, $f$, and $\nu_{\rm ps}$, with $\lambda_{0}$ the bare electron-phonon coupling parameter ($\nu_{\rm sc}=\nu_{\rm ps}=0$). We find that the enhancement of the coupling parameter $\lambda$ for $\delta <0$ strongly increases with $\nu_{\rm ps}$, and features a maximum at small SC concentration $f\sim 0.01$ in qualitative agreement with the experiments. While our model is too simple to be used to directly predict $T_c$, which also depends on, e.g., averaged SC phonon frequencies~\cite{McMillan1968,Allen1975}, it shows that $\lambda$ can be significantly modified by interactions with a strongly coupled polymer.

Our study shows that the $T_c$ of SCs can be modified by light-matter strong coupling. Just as importantly, this is made possible through a cooperative effect involving an auxiliary coupler interacting with the superconducting material. This mechanism provides both a way to change the $T_c$ and another tool to investigate the role of phonons in conventional and unconventional SCs. In particular, while the microscopic mechanisms underlying superconductivity in YBCO remain unclear~\cite{Proust2019}, our experimental results show that phonons play a role in determining the $T_c$ of this compound, in agreement with recent studies~\cite{Bernhard2000,Liu2019}. 

\begin{methods}

\subsection{Experimental methods}

The polymers, polystyrene (analytical standard for GPC; MW 200,000), polymethyl methacrylate (MW 120,000), polyvinyl acetate (MW 140,000), and toluene were purchased from ``Sigma Aldrich''. Toluene was distilled and dried before use. Silicon (Si) wafers (4 inch diameter, 500 $\mu$m-thick) purchased from ``TEDPELLA'' were cut into 2x2 cm pieces and cleaned by sonication in water followed by isopropanol. The cleaned Si wafers were dried in a hot air oven and used as substrates for thin film making. All the glass wares, spatulas, and the Pyrex tubes used for the experiments were baked at 400 \degree C for 4h, prior to the use to ensure purity. 

The superconducting powder of YBCO (YBa$_2$Cu$_3$O$_{6+\delta}$; 99.8$\%$) was purchased from ``Can Superconductors'' and used without further purification. Thin films of YBCO dispersed in different polymers were prepared as follows. The YBCO powder was well grounded and added to freshly prepared homogeneous solutions of PS (20 wt$\%$), PMMA (16 wt$\%$), PVAc (16 wt$\%$), and stirred for at least six hours at room temperature. YBCO was maintained at 10 wt$\%$ in all the polymer samples. The thin films (4 $\mu$m-thick) were prepared by spin-casting (RPM=1000, time=2minutes) the YBCO-polymer mixture onto a Si substrate for bare films, and onto a Au-coated Si substrate for strongly coupled films. Metal deposition (20 nm) was done by sputtering. The samples were stored in a vacuum box. To measure the temperature-dependent magnetization of the strongly coupled YBCO and bare films, the Si substrates containing the spin-casted films were cut into small pieces (ca. 4x4 mm), then inserted into a gelatin capsule and sealed using Kapton tape. The gelatin capsule was then placed into a plastic straw for magnetization measurements with the SQUID magnetometer.

A reported procedure was followed to prepare the powder samples of Rb$_3$C$_{60}$ SCs~\cite{Tanigaki1991}. C$_{60}$ crystals (sublimed, 99.9$\%$) and rubidium (trace metals basis, 99.6$\%$) were purchased from ``Sigma Aldrich''. A measured amount of rubidium in a glass capillary tube (0.5 mm diameter) was added to C$_{60}$ (34 mg) in a Pyrex tube (5 mm) under inert atmosphere in a glovebox. The tube containing the mixture of C$_{60}$ and rubidium was degassed with a vacuum pump and refilled with argon gas. The degassing/refilling was repeated thrice before sealing the Pyrex tube under vacuum. The sealed tube containing the precursors was heated to 400 \degree C and maintained at that temperature for 72 h. The formation of superconducting Rb$_3$C$_{60}$ was confirmed by temperature-dependent DC magnetization measurements using the SQUID magnetometer. The thin films and strongly coupled films of Rb$_3$C$_{60}$ were prepared using the powder samples. The tube containing the Rb$_3$C$_{60}$ powder was opened in a glovebox and the powder was well grounded to obtain smaller grains. It was then added to a freshly made homogeneous solution of polymers, PS (16 wt$\%$) and PMMA (16 wt$\%$) in dried toluene and stirred for 2h in the glovebox. The resulting mixture contained 6 wt$\%$ of superconducting powder relative to the polymer. The bare and strongly coupled Rb$_3$C$_{60}$ films (4 $\mu$m-thick) were prepared in the glovebox by spin-casting (RPM=1500, time=2minutes) the Rb$_3$C$_{60}$-polymer mixture onto either a Si substrate or a Au-coated Si substrate. The substrates were then cut into smaller pieces (ca. 4x4 mm) for magnetization measurements and transferred to the Pyrex tube. The Pyrex tube containing the samples was thrice degassed and refilled with argon, and subsequently sealed under vacuum. The sealed tube was then inserted into a plastic straw for magnetization measurements with the SQUID magnetometer.  

The temperature-dependent magnetization (ZFC and FC) of the samples were measured in Strasbourg using a MPMS SQUID magnetometer (``Quantum Design Europe'') at a 100 Oe magnetic field. Some samples were also measured by ``Quantum Design Europe'' in Darmstadt (Germany) using a MPMS3 SQUID magnetometer at 100 Oe, a more efficient and recent magnetometer to ensure the reproducibility of the data. The $T_c$ is determined using a fourth-order polynomial fit at low temperature and a linear fit at high temperature, and corresponds to the limit of the intersection point of the two fits when increasing the number of fitted data points in the two temperature ranges.

The infrared spectra were recorded using Bruker Vertex70 FTIR spectrometer. The dispersion curves of the strongly coupled samples were measured in ATR mode with a Variable Angle Reflection Accessory (``Bruker'' A513/Q). We placed a right-angle ZnSe prism at the sample position which was sputter-coated with 10 nm of Au and spin-coated with a thick (doped) polymer layer (5 $\mu$m) at the bottom. We then scanned the angle of incidence on the prism from 20\degree \, to 40\degree \, with steps of 0.5\degree \, for both TM and TE polarizations. The instrument response of the reflection accessory was taken into account by correcting ATR data with the reference measurement of a bare Au film sputter-coated at the bottom of the ZnSe prism.

\subsection{Theoretical methods}

The two optical phonons with bare frequencies $\omega_{\rm sc}$ and $\omega_{\rm ps}$ are assumed dispersionless and both polarized in the in-plane and out-of-plane directions. Our effective description is valid when the in-plane SPP wavelength is larger than the typical size of the superconducting crystals in the dielectric film. To lighten the notations, we denote ${\bf q}$ the in-plane wave vector and ${\bf Q}\equiv ({\bf q},q_{z})$ the 3d wave vector. Our system consisting of four coupled harmonic oscillators is described by the Hamiltonian $H_{\rm pol}=\sum_{\bf q_{\parallelsum}} H^{({\bf q})}_{\rm pol}$, where $H^{({\bf q})}_{\rm pol}$ is decomposed as $H^{({\bf q})}_{\rm pol}=H^{({\bf q})}_{\rm pt}+ H^{({\bf q})}_{\rm mat} +H^{({\bf q})}_{\rm mat-pt}$. The first term $H^{({\bf q})}_{\rm pt}=\omega_{q} a^{\dagger}_{\bf q} a_{\bf q}$ is the free photon Hamiltonian. The bosonic operator $a_{\bf q}$ ($a^{\dagger}_{\bf q}$) annihilates (creates) a photon with in-plane wave vector ${\bf q}$ and frequency $\omega_{q}$. The latter depends on the penetration depths $1/\gamma_{\rm d}$ and $1/\gamma_{\rm m}$ of the electromagnetic field in the dielectric and the metal, respectively, and is given in the Supplementary Information file. The second and third terms in $H^{({\bf q})}_{\rm pol}$ are the matter and the light-matter coupling Hamiltonians, respectively, and read 
\begin{align}
H^{({\bf q})}_{\rm mat}&=\omega_{\rm pl} \pi^{\dagger}_{\bf q}\pi_{\bf q} + \widetilde{\omega}_{\rm sc} b^{\dagger}_{\bf q}b_{\bf q}+\widetilde{\omega}_{\rm ps} p^{\dagger}_{\bf q}p_{\bf q} + \Lambda^{\rm sc-ps} \left(b_{\bf -q} + b^{\dagger}_{\bf q} \right)\left(p_{\bf q} + p^{\dagger}_{\bf -q} \right) \nonumber \\
H^{({\bf q})}_{\rm mat-pt}&=\Big[ \Omega^{\rm pl-pt}_{q} \left(\pi_{\bf -q} + \pi^{\dagger}_{\bf q} \right) + \Omega^{\rm sc-pt}_{q} \left(b_{\bf -q} + b^{\dagger}_{\bf q} \right) +  \Omega^{\rm ps-pt}_{q} \left(p_{\bf -q} + p^{\dagger}_{\bf q} \right) \Big] \Big[a_{\bf q} + a^{\dagger}_{\bf -q} \Big]. 
\label{Hpol_2D_old}
\end{align}
Here, the matter Hamiltonian is projected onto the modes coupled to light that are called bright modes, and contains the free contributions of the two phonon modes, the plasmon mode in the metal, as well as a direct dipole-dipole interaction between the two types of phonons. The bosonic operators $b_{\bf q}$ ($b^{\dagger}_{\bf q}$), $p_{\bf q}$ ($p^{\dagger}_{\bf q}$), and $\pi_{\bf q}$ ($\pi^{\dagger}_{\bf q}$) respectively annihilate (create) a SC phonon, PS phonon, and plasmon bright modes. The different coupling strengths read 
\begin{alignat*}{2}
\Lambda^{\rm sc-ps}&= \frac{\nu_{\rm sc}\nu_{\rm ps}}{2}\sqrt{\frac{f(1-f)}{\widetilde{\omega}_{\rm sc}\widetilde{\omega}_{\rm ps}}} &\quad \Omega^{\rm pl-pt}_{q}&= \sqrt{\frac{c\,\omega_{\rm pl}\left(q^{2}+\gamma^{2}_{\rm m} \right)}{\gamma_{\rm m}}} \nonumber \\
\Omega^{\rm sc- pt}_{q} &= \nu_{\rm sc} \sqrt{\frac{c f(q^2+\gamma^{2}_{\rm d})(1-e^{-\gamma_{\rm d}\ell})}{\epsilon\gamma_{\rm d}\,\widetilde{\omega}_{\rm sc}}} &\quad \Omega^{\rm ps- pt}_{q} &= \nu_{\rm ps} \sqrt{\frac{c (1-f)(q^2+\gamma^{2}_{\rm d})(1-e^{-\gamma_{\rm d}\ell})}{\epsilon\gamma_{\rm d}\,\widetilde{\omega}_{\rm ps}}},
\end{alignat*} 
with $\ell$ the quantization length of the phonon fields in the out-of-plane direction. The polariton Hamiltonian exhibits $4$ positive eigenvalues $w_{q\zeta}$ in each subspace ${\bf q}$, which can be determined using a self-consistent algorithm working as follows~\cite{Todorov2014,Hagenmuller2019}. Starting with a given polariton frequency, we use the Helmholtz equation $\epsilon_{i} (w_{q\zeta}) w_{q\zeta}^{2}/c^{2}=q^{2}-\gamma^{2}_{i}$ to determine the field penetration depths entering $H_{\rm pol}$. The dielectric functions $\epsilon_{i} (\omega)$ in the dielectric ($i={\rm d}$) and the metal ($i={\rm m}$) read $\epsilon_{\rm d}=\epsilon$ and $\epsilon_{\rm m} (\omega)=1- \omega^{2}_{\rm pl}/\omega^{2}$. The polariton Hamiltonian is then diagonalized numerically and the algorithm is repeated with the lowest eigenvalue as a new polariton frequency until convergence is reached.

The coupling between electrons in the $t_{1u}$ band of ${\rm Rb}_{3} {\rm C}_{60}$ and the on-ball phonon mode of C$_{60}$ is described by the Hamiltonian
\begin{align*}
H_{\rm el-pn}= \sum_{\bf K} \xi_{K} c^{\dagger}_{\bf K} c_{\bf K} + V \sum_{{\bf K},{\bf Q},\alpha} c^{\dagger}_{\bf K} c_{{\bf K}-{\bf Q}} \left(S_{{\bf Q}\alpha}+ S^{\dagger}_{{\bf -Q}\alpha}\right),  
\end{align*}
with the 3d SC phonon annihilation and creation operators $S_{{\bf Q}\alpha}$ and $S^{\dagger}_{{\bf Q}\alpha}$. The latter are related to the quasi-2d bright phonon modes by $b_{\bf q}\equiv \sum_{q_{z},\alpha} f_{\alpha} ({\bf Q}) S_{{\bf Q}\alpha}$, where $f_{\alpha} ({\bf Q})$ stems from the overlap between the electric field and the SC phonon field in the out-of-plane direction $z$, and is given in the Supplementary Information file. The matrix element $V$ does not depend on the wave vector ${\bf Q}$ at the lowest (tight-binding) order, as shown in Ref.~[\onlinecite{Lannoo1991}]. The fermionic operator $c_{\bf K}$ ($c^{\dagger}_{\bf K}$) annihilates (creates) a spin-less electron with 3d wave vector ${\bf K}$ and energy $\xi_{K}$ relative to the Fermi energy. For simplicity, we assume a spherical Fermi surface providing $\xi_{K} =\hbar^{2} (K^{2}- K^{2}_{\rm F})/(2 m)$, with $m$ the electron effective mass and $K_{\rm F}$ the Fermi wave vector. At zero temperature, the electron-phonon dimensionless coupling parameter $\lambda$ reads 
\begin{align*}
\lambda= \frac{1}{N(0)}\sum_{\bf K} \delta (\xi_{K}) \Re \left(-\partial_{\omega} \overline{\Sigma}_{\bf K} (\omega) \vert_{\omega=0} \right),
\end{align*}
where $\Re$ stands for real part, $N(0)$ is the electron density of states at the Fermi level, and $\partial_{\omega} \overline{\Sigma}_{\bf K} (\omega) \vert_{\omega=0}$ denotes the frequency derivative of the retarded electron self-energy $\overline{\Sigma}_{\bf K} (\omega)$ evaluated at $\omega=0$. We compute this self-energy using the polariton Hamiltonian Eq.~(\ref{Hpol_2D_old}), and then derive the expression of the parameter $\lambda$. The relative enhancement of $\lambda$ is obtained as
\begin{align*}
\frac{\Delta\lambda}{\lambda_{0}}= \frac{1}{2N^{2}(0)} \sum_{\bf K} \delta (\xi_{K}) \sum_{{\bf Q},\alpha} \left(\varphi_{q} - 1 \right) \vert f_{\alpha} ({\bf Q}) \vert^{2} \delta (\xi_{{\bf K}-{\bf Q}}),
\end{align*}
where the function $\varphi_{q}$ describes the renormalization of the SC phonon energy due to the coupling to PS phonons and photons, and is provided in the Supplementary Information file. Due to the breaking of translational invariance in the out-of-plane direction (Au layer), $\Delta \lambda\propto 1/(K_{\rm F} \ell)$ is found to be proportional to the ratio between the 2d and the 3d electron density of states. Here, we choose $\ell = 20 {\rm nm}$ of the same order of magnitude than the penetration depth of SPPs in the dielectric film for $q\sim 2 K_{\rm F}$, at the border of the validity domain of our 3d model ($K_{\rm F}\ell = 1$). We find that the enhancement of the electron-phonon coupling reaches its maximum value for $K_{\rm F}\ell \ll 1$, which corresponds to a 2d configuration. The diameter of the Fermi surface $2 K_{\rm F}=0.1 {\rm nm}^{-1}$ in the calculations. 
 
\end{methods}

\bibliographystyle{naturemag}
\bibliography{supra_light_main}

\begin{addendum}
\item The authors acknowledge support from the International Center for Frontier Research in Chemistry (icFRC, Strasbourg), the ANR Equipex Union (ANR-10-EQPX-52-01), the Labex NIE projects (ANR-11-LABX-0058 NIE) and CSC (ANR-10-LABX-0026 CSC) within the ``Investissement d'Avenir'' program ANR-10-IDEX-0002-02, the ERC (project no 788482 MOLUSC), and ANR ``ERA-NET QuantERA'' projet ``RouTe'' (ANR-18-QUAN-0005-01). M.S. acknowledges funding from the European Union's Horizon 2020 programme under the Marie Sk\l{}odowska-Curie grant agreement no 753228. G.P. acknowledges additional support from the ``Institut Universitaire de France'' (IUF), and the University of Strasbourg Institute for Advanced Studies (USIAS). D.~H. acknowledges valuable discussions with Y.~Laplace. The authors thank T.~Adler and S.~Riesner from ``Quantum Design Europe'' for measurements in Darmstadt.  
%\clearpage
\item[Author contributions] T.~W.~E. initiated the study and oversaw the experiments. Experiments were performed by A.~T., E.~D., K.~N., T.~C., and M.~S.   Theoretical investigations were led by D.~H. and G.~P. and conducted by D.~H., S.~S., J.~S., and G.~P. The manuscript was written by A.~T., D.~H., S.~S., J.~S., G.~P., and T.~W.~E.; all authors discussed the results and contributed to the manuscript.
\item[Competing Interests] The authors declare that they have no competing financial interests.
\item[Correspondence] Correspondence and requests for materials should be addressed to T.~W.~Ebbesen~(email: ebbesen@unistra.fr) and G. Pupillo~(email: pupillo@unistra.fr).
\end{addendum}

\end{document}

\title{Supplementary information for ``Exploring Superconductivity under Strong Coupling with
the Vacuum Electromagnetic Field''}

\author{A.~Thomas}
\affiliation{ISIS (UMR 7006) \& icFRC, University of Strasbourg and CNRS, 8 all\'ee G. Monge, 67000 Strasbourg, France}

\author{E.~Devaux}
\affiliation{ISIS (UMR 7006) \& icFRC, University of Strasbourg and CNRS, 8 all\'ee G. Monge, 67000 Strasbourg, France}

\author{K.~Nagarajan}
\affiliation{ISIS (UMR 7006) \& icFRC, University of Strasbourg and CNRS, 8 all\'ee G. Monge, 67000 Strasbourg, France}

\author{T.~Chervy}
\affiliation{ISIS (UMR 7006) \& icFRC, University of Strasbourg and CNRS, 8 all\'ee G. Monge, 67000 Strasbourg, France}

\author{M.~Seidel}
\affiliation{ISIS (UMR 7006) \& icFRC, University of Strasbourg and CNRS, 8 all\'ee G. Monge, 67000 Strasbourg, France}

\author{D.~Hagenm\"uller}
\affiliation{ISIS (UMR 7006) \& icFRC, University of Strasbourg and CNRS, 8 all\'ee G. Monge, 67000 Strasbourg, France}

\author{S.~Sch\"utz}
\affiliation{ISIS (UMR 7006) \& icFRC, University of Strasbourg and CNRS, 8 all\'ee G. Monge, 67000 Strasbourg, France}

\author{J.~Schachenmayer}
\affiliation{ISIS (UMR 7006) \& icFRC, University of Strasbourg and CNRS, 8 all\'ee G. Monge, 67000 Strasbourg, France}

\author{C.~Genet}
\affiliation{ISIS (UMR 7006) \& icFRC, University of Strasbourg and CNRS, 8 all\'ee G. Monge, 67000 Strasbourg, France}

\author{G.~Pupillo}
\affiliation{ISIS (UMR 7006) \& icFRC, University of Strasbourg and CNRS, 8 all\'ee G. Monge, 67000 Strasbourg, France}

\author{T.~W.~Ebbesen}
\affiliation{ISIS (UMR 7006) \& icFRC, University of Strasbourg and CNRS, 8 all\'ee G. Monge, 67000 Strasbourg, France}

\maketitle 

\onecolumngrid

\date{}

%\tableofcontents

\section{Extended data}
\label{ext_data}

In this section, we provide extended data obtained from temperature-dependent magnetization measurements and Fourier-transform infrared spectroscopy (FT-IR). The grey solid lines in Figs.~\ref{fig_Ftir}\textbf{b}, \ref{fig_magn}, and \ref{fig_fit_YBCO} correspond to a fourth-order polynomial fit at low temperature and a linear fit at high temperature. The $T_c$ corresponds to the limit of the intersection point of the two fits when increasing the number of fitted data points in the two temperature ranges.

%The FT-IR transmission spectra of polyvinyl acetate (PVAc) and YBCO are displayed in Fig.~\ref{fig_Ftir} together with magnetization measurements of a YBCO powder dispersed in PVAc on Au. Magnetization measurements in the zero-field-cooled (ZFC) and field-cooled (FC) modes are shown in Figs.~\ref{fig_magn}, \ref{fig_ps}, and \ref{fig_Rb}, for YBCO dispersed in polystyrene (PS) on Ag and Pt, a PS film on an Au-coated Si substrate, and a Rb$_3$C$_{60}$ powder, respectively.

\begin{figure}[ht]
\centerline{\includegraphics[width=0.4\columnwidth]{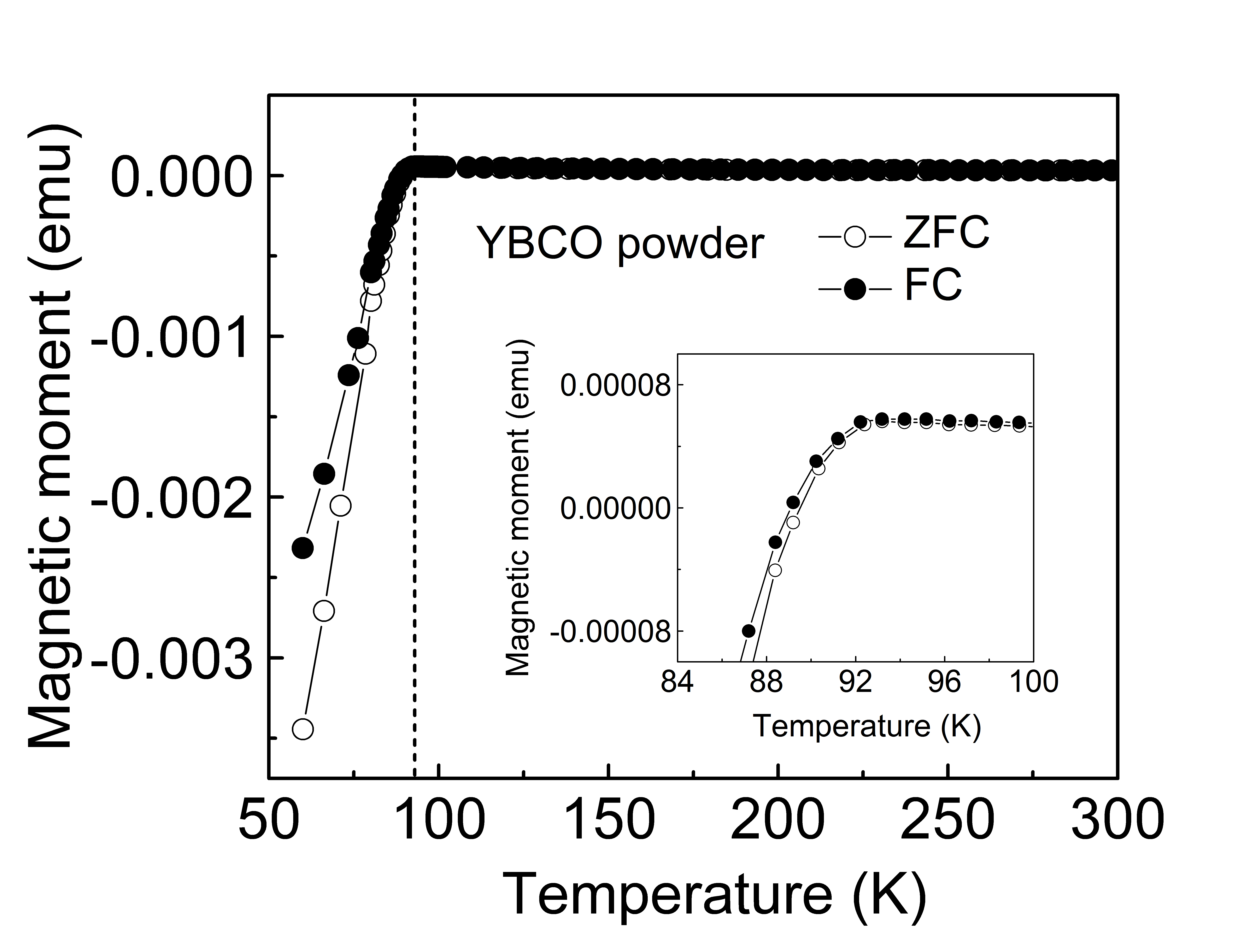}}
\caption{Magnetization measurements of the YBCO powder. Temperature-dependent magnetization of YBCO powder in the temperature range of 50 K - 300 K for ZFC and FC at 50 Oe field. The inset shows a zoom of the magnetization data providing a $T_c$ of 92 K.} 
\label{fig_YBCO_alone}
\end{figure}

\begin{figure}[ht]
\centerline{\includegraphics[width=0.8\columnwidth]{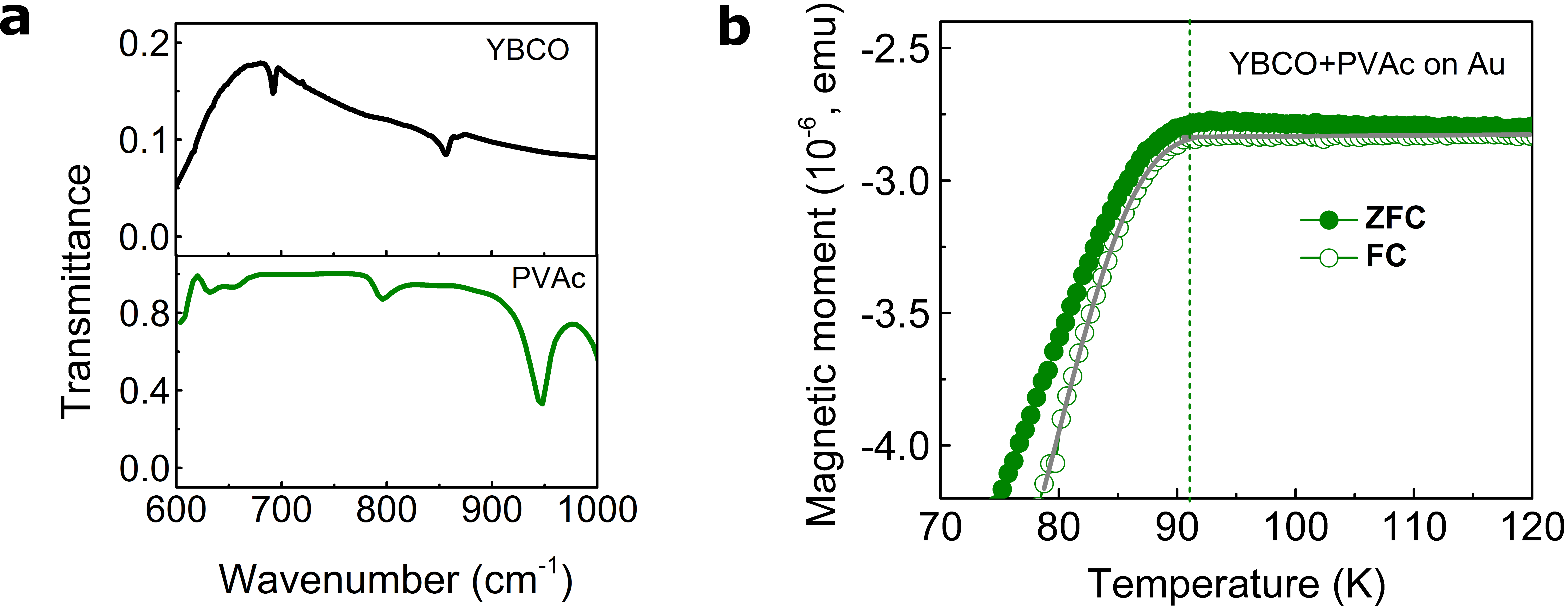}}
\caption{Control experiments with off-resonant YBCO. \textbf{a}, FT-IR transmission spectrum of PVAc (green) showing the absence of strong IR vibrations in the region of YBCO (black) vibrational modes. \textbf{b}, Temperature dependent magnetization for off-resonant YBCO films (YBCO+PVAc on Au) in the zero-field-cooled (ZFC) and 100 Oe field-cooled (FC) modes. The dashed line indicates the $T_c$ (92 K) showing the absence of any strong coupling effect.} 
\label{fig_Ftir}
\end{figure}

\begin{figure}[ht]
\centerline{\includegraphics[width=0.8\columnwidth]{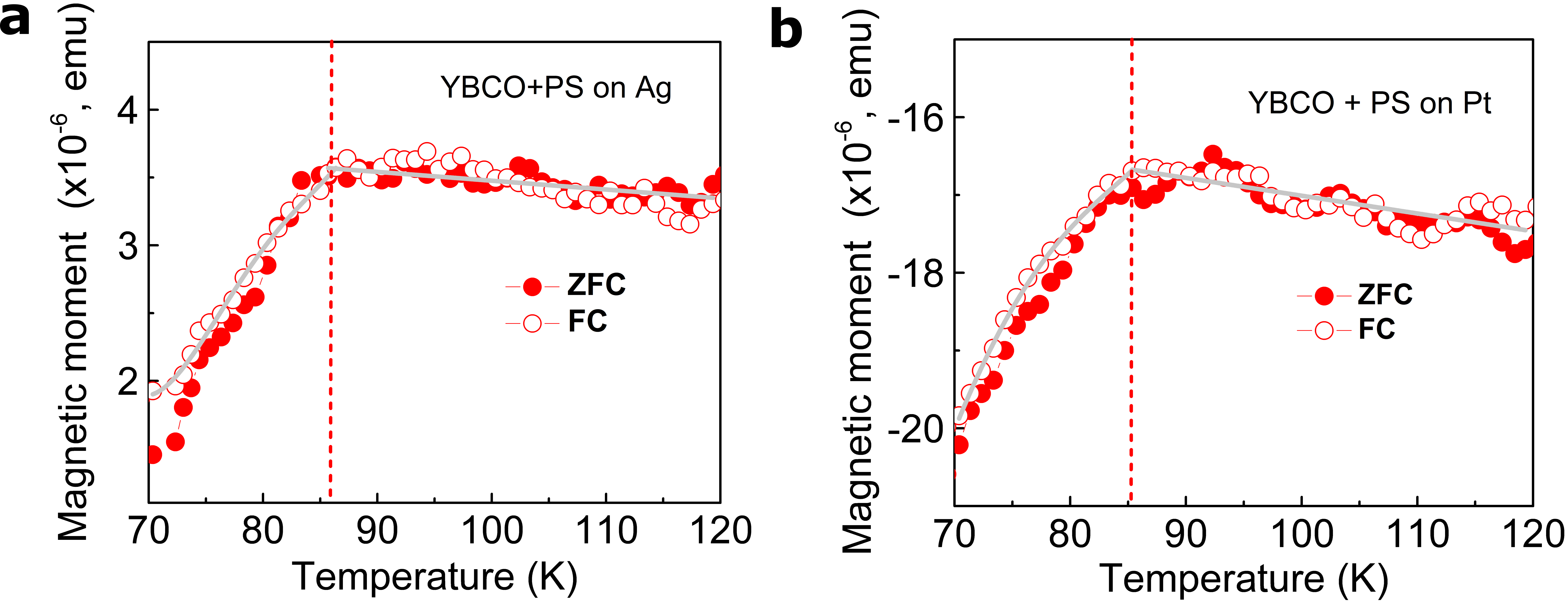}}
\caption{Magnetic susceptibility of strongly coupled YBCO on Ag and Pt. Temperature dependent magnetization of YBCO films strongly coupled to the SPP modes of Ag (\textbf{a}) and Pt (\textbf{b}) in the zero-field-cooled (ZFC) and 100 Oe field-cooled (FC) modes. Dashed lines indicate the $T_c$ (86 K).} 
\label{fig_magn}
\end{figure}

\begin{figure}[ht]
\centerline{\includegraphics[width=0.35\columnwidth]{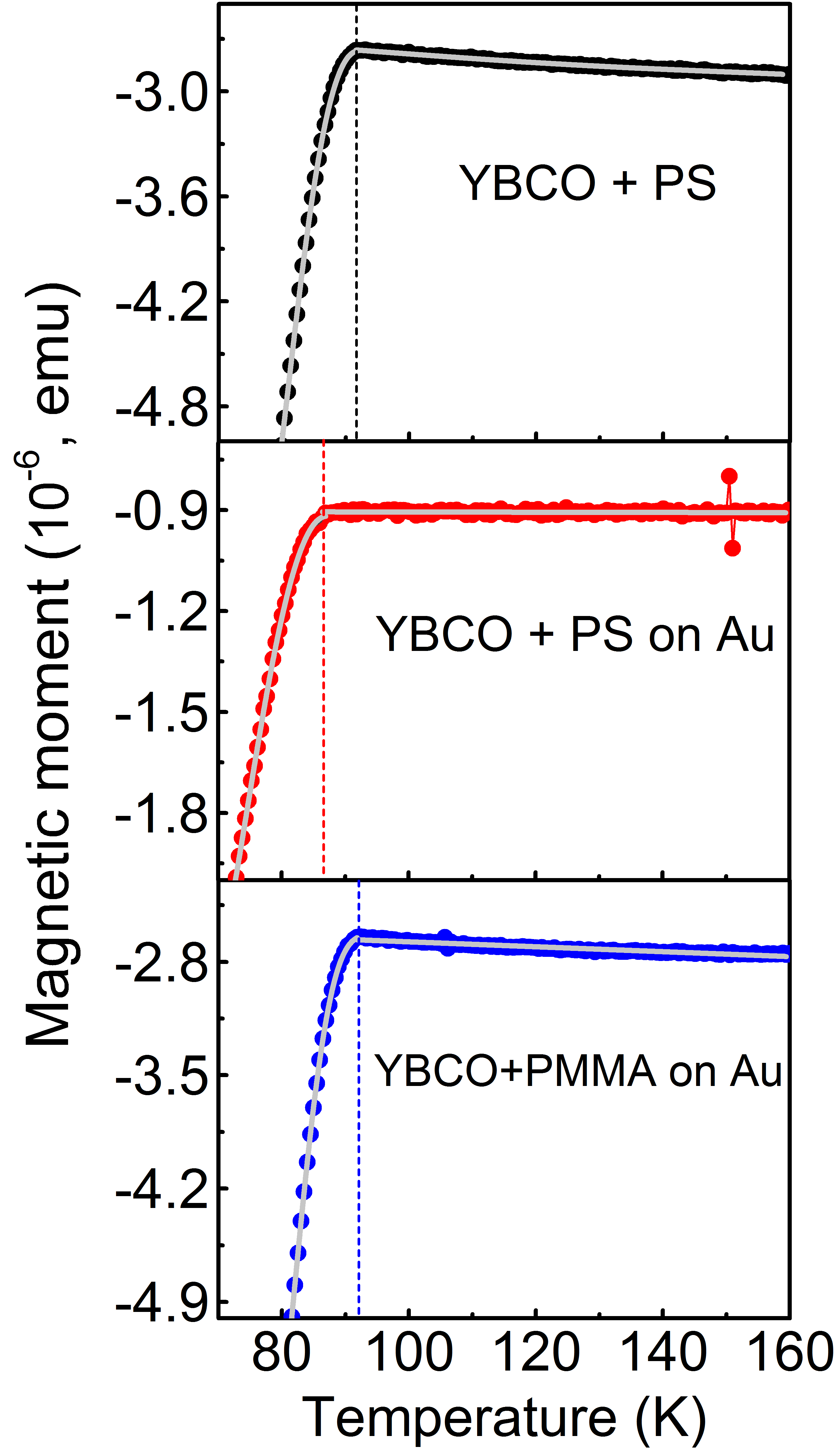}}
\caption{Onset of superconductivity for YBCO samples. Field-cooled temperature-dependent magnetization at 100 Oe between 70 K and 160 K of a bare film of YBCO+PS on Si (top), strongly coupled YBCO+PS on Au (middle), and cooperatively off-resonant YBCO+PMMA on Au (bottom).} 
\label{fig_fit_YBCO}
\end{figure}

\begin{figure}[ht]
\centerline{\includegraphics[width=0.8\columnwidth]{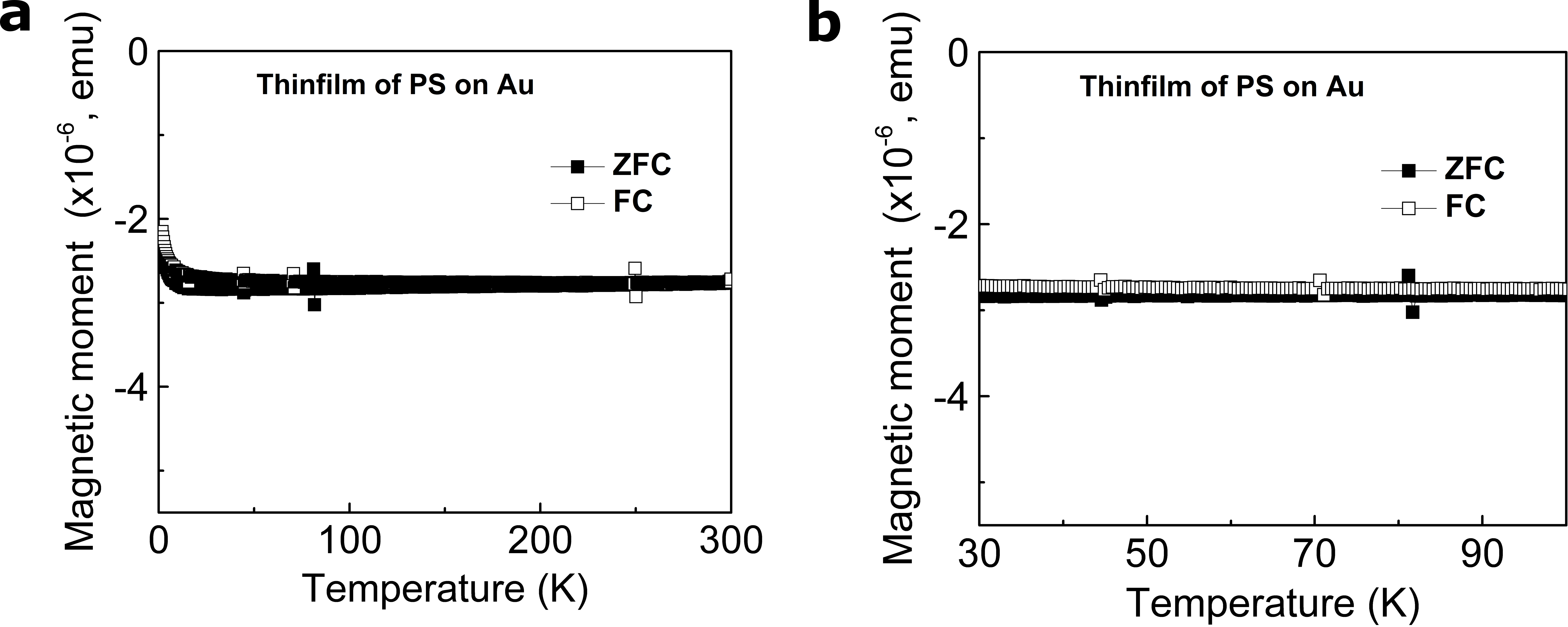}}
\caption{Magnetization measurements of PS films spin-coated on Au. \textbf{a}, Temperature dependent magnetization of spin-coated PS films (4 micron thick) on an Au-coated Si substrate showing the absence of any superconducting properties in the temperature range 0 K -300 K in the zero-field-cooled (ZFC) and 100 Oe field-cooled (FC) modes. \textbf{b}, Magnetization data magnified in the region of interest for the present study.} 
\label{fig_ps}
\end{figure}

\begin{figure}[ht]
\centerline{\includegraphics[width=0.5\columnwidth]{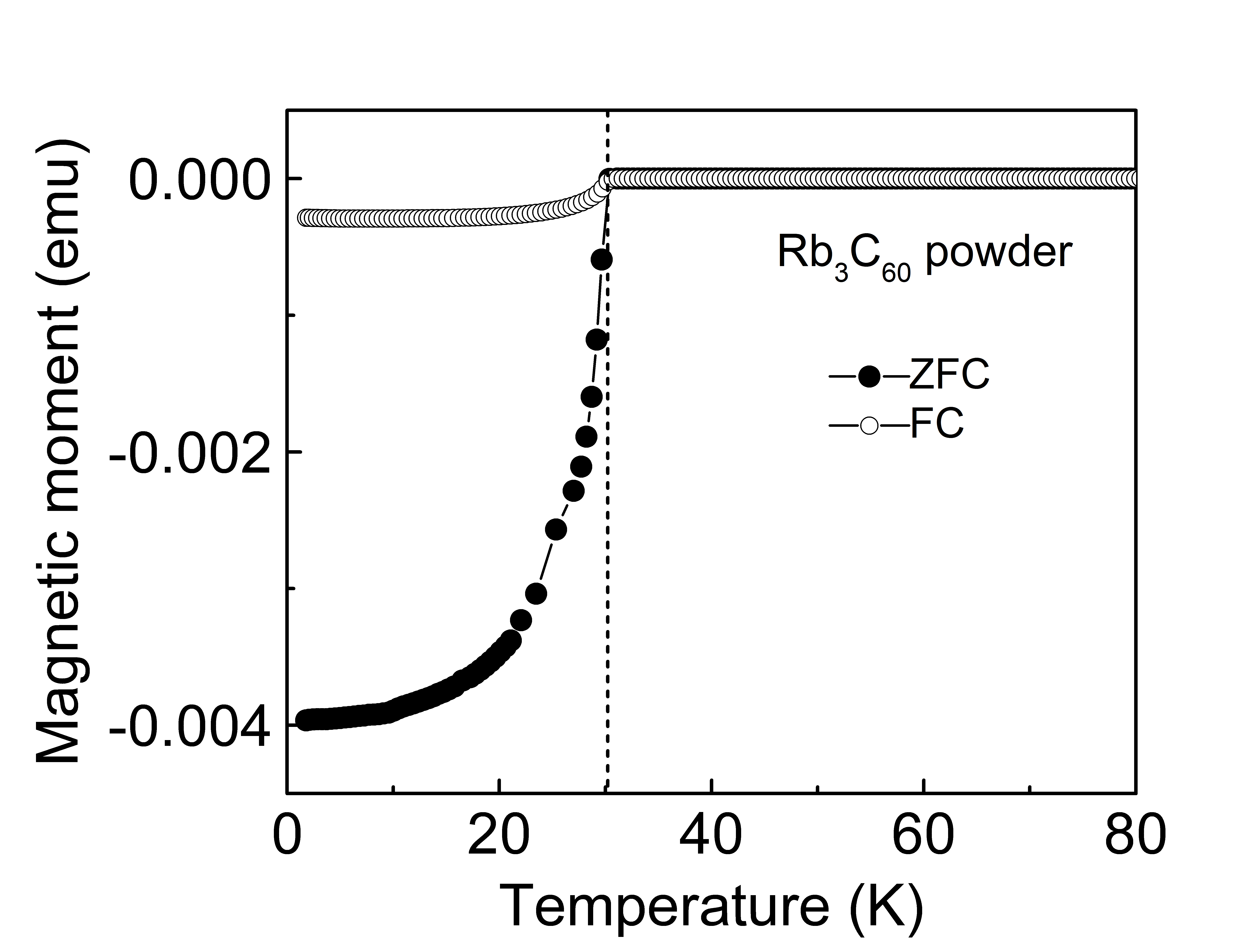}}
\caption{Magnetization measurements of the Rb$_3$C$_{60}$ powder. Temperature-dependent magnetization of the Rb$_3$C$_{60}$ powder in the temperature range of 0 K - 80 K in the zero-field-cooled (ZFC) and 20 Oe field-cooled (FC) modes.} 
\label{fig_Rb}
\end{figure}

\begin{figure}[ht]
\centerline{\includegraphics[width=0.8\columnwidth]{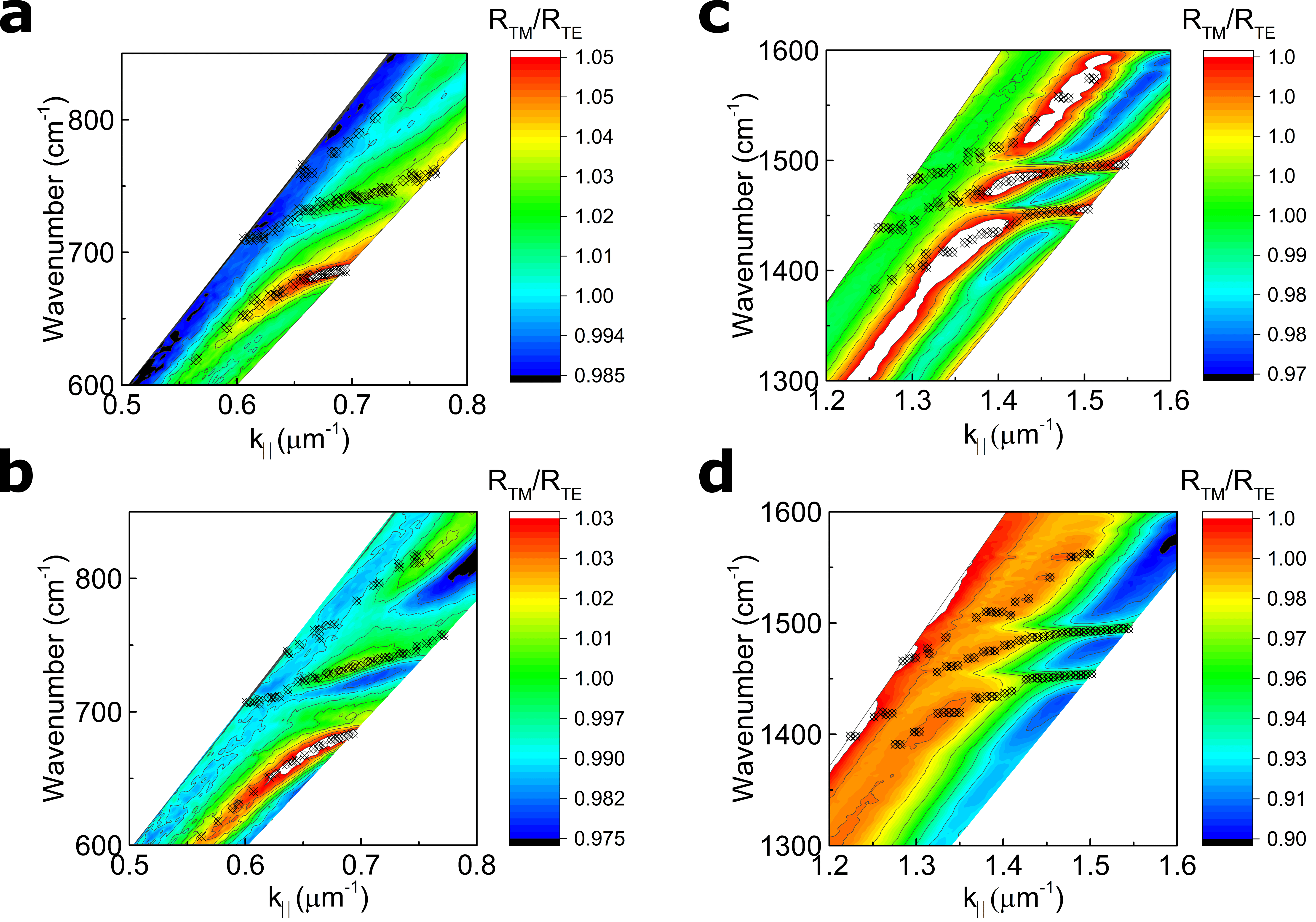}}
\caption{Dispersion of the films as a function of the in-plane wave vector $k_{\parallelsum}$ (parallel to the Au layer). \textbf{a} \& \textbf{c} correspond to the bare films (PS on Au) in the low and high-frequency ranges, respectively. \textbf{b} \& \textbf{d} correspond to the superconducting films (PS + superconductor on Au) with YBCO (\textbf{b}) and Rb$_3$C$_{60}$ (\textbf{d}). In order to correct for residual absorption in transmission, e.g., on top of the ZnSe prism, we measured the ratio R$_{\rm TM}$/R$_{\rm TE}$ of the reflectivities for TM and TE polarizations.} 
\label{fig_disp}
\end{figure}

\clearpage

\section{Theoretical model}
\label{theory_mod}

Here we introduce the theoretical model providing a qualitative explanation for a cooperative
enhancement of the electron-phonon coupling when phonons are strongly coupled to SPPs. The different parts of the Hamiltonian are derived in Secs.~\ref{mat_HH}, \ref{pho_H}, and \ref{light_ma}. In Sec.~\ref{popo_H}, we provide the diagonalization procedure of this Hamiltonian and discuss the hybridization mechanism between the vibrational modes of the superconductor (SC) and polystyrene (PS). In Sec.~\ref{elec_pho}, we derive the electron-phonon coupling parameter $\lambda$ and explain how the dressing of SC phonons by surface plasmon polaritons (SPPs) and the vibrational modes of PS can lead to an enhancement of $\lambda$. In Sec.~\ref{fit_sys}, we use our model to fit the dispersion curves obtained from the attenuated total reflexion measurements in order to extract the typical coupling strengths of PS to light.

The film containing the SC dispersed in PS is described as an effective, homogeneous dielectric medium with background dielectric constant of the polymer $\epsilon$, and contains two phonon modes with frequencies $\omega_{\rm sc}$ and $\omega_{\rm ps}$ corresponding to an on-ball mode of C$_{60}$ and the quasi-resonant mode of PS, respectively. These two phonons are assumed dispersionless, both polarized in the out-of-plane and in-plane directions, and interact with SPPs generated by the metal-dielectric interface. We call $z$ the out-of-plane axis perpendicular to the metal-dielectric interface (unit vector ${\bf u}_z$), while ${\bf u}_{\parallelsum}$ denotes the unit vector parallel to the interface. The dielectric film is located in the region $z>0$ and the metal (Au) in the region $z<0$. Denoting as $q$ the in-plane wave vector, our effective description is valid when the SPP wavelength $2\pi/q$ is much larger than the typical size of the superconducting inclusions in the film. The Au layer exhibits an electronic plasmon mode with plasma frequency $\omega_{\rm pl}$. The total Hamiltonian is decomposed as $H = H_{\rm pol} + H_{\rm el-pn}$, where $H_{\rm pol}$ denotes the polariton Hamiltonian and $H_{\rm el-pn}$ is the electron-phonon coupling Hamiltonian in the SC. The polariton Hamiltonian is derived in the Power-Zienau-Woolley representation~\cite{Babiker439,todorovY} and reads $H=H_{\rm mat}+H_{\rm pt}+H_{\rm mat-pt}$. The first term describes the matter part, the second one the photonic part, and the third one the light-matter coupling. 

\subsection{Matter Hamiltonian}
\label{mat_HH}

The matter Hamiltonian $H_{\rm mat}$ is decomposed as $H_{\rm mat}=H_{\rm pl}+H_{\rm sc}+H_{\rm ps}+H_{\rm P^{2}}$, where $H_{\rm pl}=\omega_{\rm pl}\sum_{{\bf Q},\alpha} \Pi^{\dagger}_{{\bf Q},\alpha}\Pi_{{\bf Q},\alpha}$, $H_{\rm sc}=\omega_{\rm sc}\sum_{{\bf Q},\alpha} S^{\dagger}_{{\bf Q},\alpha}S_{{\bf Q},\alpha}$ and $H_{\rm ps}=\omega_{\rm ps}\sum_{{\bf Q},\alpha} P^{\dagger}_{{\bf Q},\alpha}P_{{\bf Q},\alpha}$ denote the free plasmon, SC phonon, and PS phonon contributions, respectively. The bosonic operators $S_{{\bf Q}\alpha}$ ($S^{\dagger}_{{\bf Q}\alpha}$), $P_{{\bf Q}\alpha}$ ($P^{\dagger}_{{\bf Q}\alpha}$) and $\Pi_{{\bf Q}\alpha}$ ($\Pi^{\dagger}_{{\bf Q}\alpha}$) respectively annihilate (create) a phonon in the SC material, a phonon in the polymer matrix, and a plasmon in the metal with 3d wave vector ${\bf Q}\equiv ({\bf q},q_{z})$ and polarization along the direction ${\bf u}_{\alpha}$ ($\alpha=z,\parallelsum$). Calling ${\bf R}\equiv ({\bf r},z)$ the 3d position vector, the term $H_{\rm P^{2}}=\frac{1}{2\epsilon_{0}\epsilon}\int \!d{\bf R} \, {\bf P}^{2} ({\bf R})$ is proportional to the square polarization density field ${\bf P}$ in the dielectric film, which can be decomposed as ${\bf P}={\bf P}_{\rm sc}+ {\bf P}_{\rm ps}$ with 
\begin{align}
{\bf P}_{\rm sc} ({\bf R})&= \nu_{\rm sc}\sqrt{\frac{\hbar \epsilon_{0}\epsilon f}{2\omega_{\rm sc} S \ell_{\rm d}}} \sum_{{\bf Q},\alpha} \left(S_{{\bf -Q}\alpha} + S^{\dagger}_{{\bf Q}\alpha}\right) {\bf u}_{\alpha} e^{-\mi {\bf Q}\cdot {\bf R}} \Theta (z)\nonumber \\
{\bf P}_{\rm ps} ({\bf R})&= \nu_{\rm ps}\sqrt{\frac{\hbar \epsilon_{0}\epsilon (1-f)}{2\omega_{\rm ps} S \ell_{\rm d}}} \sum_{{\bf Q},\alpha} \left(P_{{\bf -Q}\alpha} + P^{\dagger}_{{\bf Q}\alpha}\right) {\bf u}_{\alpha} e^{-\mi {\bf Q}\cdot {\bf R}} \Theta (z).
\label{pola}
\end{align}
Here, $\Theta(z)$ is the heaviside function, $\epsilon_0$ the vacuum permittivity, $\ell_{\rm d}$ is the quantization length of the phonon fields in $z$ direction, $S$ the surface of the film, and $f$ the SC filling fraction in the film. The parameters $\nu_{\rm sc}$ and $\nu_{\rm ps}$ denote the ionic plasma frequencies~\cite{hagen} associated to the SC and PS phonon dipole moments, respectively. In the Hamiltonian $H_{\rm P^{2}}$, the two self-interaction terms $\propto {\bf P}^{2}_{\rm sc}$ and $\propto {\bf P}^{2}_{\rm ps}$ lead to a depolarization shift of the bare phonon frequencies: $\widetilde{\omega}_{\rm sc}=\sqrt{\omega^{2}_{\rm sc}+\nu^{2}_{\rm sc}f}$ and $\widetilde{\omega}_{\rm ps}=\sqrt{\omega^{2}_{\rm ps}+\nu^{2}_{\rm ps}(1-f)}$. The term $\propto {\bf P}_{\rm sc} \cdot {\bf P}_{\rm ps}$ corresponds to a direct dipole-dipole interaction between the two types of phonons and leads to phonon hybridization in the film. Using Eq.~(\ref{pola}), the matter Hamiltonian can be written as
\begin{align}
H_{\rm mat}=\omega_{\rm pl}\sum_{{\bf Q},\alpha} \Pi^{\dagger}_{{\bf Q}\alpha}\Pi_{{\bf Q}\alpha}+\widetilde{\omega}_{\rm sc}\sum_{{\bf Q},\alpha} \widetilde{S}^{\dagger}_{{\bf Q}\alpha}\widetilde{S}_{{\bf Q}\alpha}+\widetilde{\omega}_{\rm ps}\sum_{{\bf Q},\alpha} \widetilde{P}^{\dagger}_{{\bf Q}\alpha}\widetilde{P}_{{\bf Q}\alpha} + \Lambda^{\rm sc-ps} \sum_{{\bf Q},\alpha} \left(\widetilde{S}_{{\bf -Q}\alpha} + \widetilde{S}^{\dagger}_{{\bf Q}\alpha} \right)\left(\widetilde{P}_{{\bf Q}\alpha} + \widetilde{P}^{\dagger}_{{\bf -Q}\alpha} \right),
\label{mat_tot}
\end{align}
with the dipole-dipole coupling strength $\Lambda^{\rm sc-ps}= \frac{\nu_{\rm sc}\nu_{\rm ps}}{2} \sqrt{\frac{f(1-f)}{\widetilde{\omega}_{\rm sc}\widetilde{\omega}_{\rm ps}}}$, and where the operators
\begin{align*}
\widetilde{S}_{{\bf Q}\alpha} =\frac{\widetilde{\omega}_{\rm sc}+\omega_{\rm sc}}{2\sqrt{\widetilde{\omega}_{\rm sc}\omega_{\rm sc}}} S_{{\bf Q}\alpha} + \frac{\widetilde{\omega}_{\rm sc}-\omega_{\rm sc}}{2\sqrt{\widetilde{\omega}_{\rm sc}\omega_{\rm sc}}} S^{\dagger}_{{\bf -Q}\alpha} \qquad \widetilde{S}^{\dagger}_{{\bf -Q}\alpha} = \frac{\widetilde{\omega}_{\rm sc}-\omega_{\rm sc}}{2\sqrt{\widetilde{\omega}_{\rm sc}\omega_{\rm sc}}} S_{{\bf Q}\alpha} + \frac{\widetilde{\omega}_{\rm sc}+\omega_{\rm sc}}{2\sqrt{\widetilde{\omega}_{\rm sc}\omega_{\rm sc}}} S^{\dagger}_{{\bf -Q}\alpha}
\end{align*}
allow to write the sum of the free contribution $\omega_{\rm sc}\sum_{{\bf Q},\alpha} S^{\dagger}_{{\bf Q},\alpha}S_{{\bf Q},\alpha}$ and the self-interaction term $\propto {\bf P}^{2}_{\rm sc}$ in diagonal form. The same definitions hold for $\widetilde{P}_{{\bf Q}\alpha}$ and $\widetilde{P}^{\dagger}_{{\bf Q}\alpha}$ with the replacements $\omega_{\rm sc} \to \omega_{\rm ps}$ and $\widetilde{\omega}_{\rm sc} \to \widetilde{\omega}_{\rm ps}$. It is convenient to fully diagonalize the phonon part of the Hamiltonian Eq.~(\ref{mat_tot}) using the hybrid phonon modes $\Gamma_{{\bf Q}\alpha j}=X_{j} S_{{\bf Q}\alpha} + \widetilde{X}_{j} S^{\dagger}_{{\bf -Q}\alpha} + Y_{j} P_{{\bf Q}\alpha} + \widetilde{Y}_{j} P^{\dagger}_{{\bf -Q}\alpha}$. Solving the system $[\Gamma_{{\bf Q}\alpha j},H_{\rm mat}]=\omega_{j} \Gamma_{{\bf Q}\alpha j}$, one can write the matter Hamiltonian as
\begin{align*}
H_{\rm mat}=\omega_{\rm pl}\sum_{{\bf Q},\alpha} \Pi^{\dagger}_{{\bf Q},\alpha}\Pi_{{\bf Q},\alpha}+\sum_{{\bf Q},\alpha,j} \omega_{j}  \Gamma^{\dagger}_{{\bf Q}\alpha j}\Gamma_{{\bf Q}\alpha j},
\end{align*}
with 
\begin{align}
2\omega^{2}_{j}= \widetilde{\omega}^{2}_{\rm sc} + \widetilde{\omega}^{2}_{\rm ps} \pm \sqrt{\left(\widetilde{\omega}^{2}_{\rm sc} - \widetilde{\omega}^{2}_{\rm ps} \right)^{2}+ 4 \nu^{2}_{\rm sc} \nu^{2}_{\rm ps} f (1-f)} \qquad j=\pm.
\label{E_hyb}
\end{align}
The hybrid mode frequencies $\omega_{j}$ are shown in Fig.~\ref{fig_hopfield}\textbf{a} as a function of the detuning $\delta=\omega_{\rm ps}-\omega_{\rm sc}$. For $f=0.02$, $\nu_{\rm sc}=0.01 \omega_{\rm sc}$, and $\nu_{\rm ps}=0.2 \omega_{\rm sc}$, an anticrossing occurs at $\delta\approx -0.02\omega_{\rm sc}$ (hardly visible in the figure). The latter corresponds to the condition $\widetilde{\omega}_{\rm ps}=\widetilde{\omega}_{\rm sc}$, where hybridization between SC and PS phonons is maximum ($50\%-50\%$). For $\nu_{\rm sc} \ll \nu_{\rm ps}\ll \omega_{\rm sc}$, $\delta \ll \omega_{\rm sc}$, and using $\widetilde{\omega}_{\rm sc}=\sqrt{\omega^{2}_{\rm sc}+\nu^{2}_{\rm sc}f}$ and $\widetilde{\omega}_{\rm ps}=\sqrt{\omega^{2}_{\rm ps}+\nu^{2}_{\rm ps}(1-f)}$, the resonance condition $\widetilde{\omega}_{\rm ps}=\widetilde{\omega}_{\rm sc}$ provides 
$$f=1+\frac{2\delta}{\omega_{\rm sc}}\left(1+\frac{\delta}{\omega_{\rm sc}}\right)^{2}\left(\frac{\omega_{\rm sc}}{\nu_{\rm ps}}\right)^{2}$$ and $$\frac{\nu_{\rm ps}}{\omega_{\rm sc}}=\sqrt{\frac{2}{1-f}\left(1+\frac{\delta}{\omega_{\rm sc}}\right)^{2} \frac{\vert\delta\vert}{\omega_{\rm sc}}},$$ which correspond to the dashed lines in Fig.~4\textbf{c} of the main text.

\subsection{Photon Hamiltonian}
\label{pho_H}

The free photon Hamiltonian reads $H_{\rm pt} = \frac{1}{2\epsilon_{0}}\int \!d{\bf R} \, \frac{{\bf D}^{2}({\bf R})}{\epsilon (z)}+ \frac{1}{2\epsilon_{0}c^{2}}\int \!d{\bf R}\, {\bf H}^{2}({\bf R})$, where $\epsilon (z) = \epsilon$ for $z>0$ and $\epsilon (z) = 1$ for $z<0$. The different components of the displacement and magnetic fields ${\bf D}$ and ${\bf H}$ read~\cite{todorovY}
\begin{align}
D_{z} ({\bf R})&= \sum_{\bf q} \mi q \sqrt{\frac{\epsilon_{0}\hbar \omega_{q}}{\alpha_{q}S}} \left(a_{\bf q} + a^{\dagger}_{\bf -q} \right) e^{\mi {\bf q}\cdot {\bf r}} \left(\Theta (z) e^{-\gamma_{\rm d} z} + \Theta (-z) e^{\gamma_{\rm m} z} \right) \nonumber \\
D_{\parallelsum} ({\bf R})&= \sum_{\bf q} \sqrt{\frac{\epsilon_{0}\hbar \omega_{q}}{\alpha_{q}S}} \left(a_{\bf q} + a^{\dagger}_{\bf -q} \right) e^{\mi {\bf q}\cdot {\bf r}} \left(\Theta (z) \gamma_{\rm d} e^{-\gamma_{\rm d} z} - \Theta (-z) \gamma_{\rm m} e^{\gamma_{\rm m} z} \right) \nonumber \\
H_{\perp} ({\bf R})&= \sum_{\bf q} \mi \omega_{q} \sqrt{\frac{\epsilon_{0}\hbar \omega_{q}}{\alpha_{q}S}} \left(a_{\bf q} - a^{\dagger}_{\bf -q} \right) e^{\mi {\bf q}\cdot {\bf r}} \left(\Theta (z) e^{-\gamma_{\rm d} z} + \Theta (-z) e^{\gamma_{\rm m} z} \right),
\label{field}
\end{align}
where $a_{\bf q}$ and $a^{\dagger}_{\bf q}$ are the bosonic annihilation and creation operators of a photon with in-plane wave vector ${\bf q}$, and $1/\gamma_{\rm d}$ and $1/\gamma_{\rm m}$ denote the penetration depths of the surface waves in the dielectric film and metal layer, respectively. Using Eq.~(\ref{field}), the photon Hamiltonian can be put in the usual form $H_{\rm pt}= \sum_{\bf q} \omega_{q} a^{\dagger}_{\bf q}a_{\bf q}$ with the photon frequency $\omega_{q}=c \sqrt{\alpha_{q}/\beta_{q}}$ and 
\begin{align*}
&\alpha_{q}=\frac{q^{2}+\gamma^{2}_{\rm d}}{\epsilon \gamma_{\rm d}} \left(1-e^{-2 \gamma_{\rm d} \ell_{\rm d}} \right) + \frac{q^{2}+\gamma^{2}_{\rm m}}{\gamma_{\rm m}} \qquad \qquad
\beta_{q}=\frac{1-e^{-2 \gamma_{\rm d} \ell_{\rm d}}}{\gamma_{\rm d}} + \frac{1}{\gamma_{\rm m}}.&
\end{align*}

\subsection{Light-matter coupling}
\label{light_ma}

The light-matter coupling Hamiltonian reads $H_{\rm mat-pt}=-\frac{1}{\epsilon_{0}}\int \!d{\bf R} \, {\bf P} ({\bf R}) \cdot \frac{{\bf D} ({\bf R})}{\epsilon (z)}$. Using Eqs.~(\ref{pola}), (\ref{field}), as well as the plasmon polarization field in the metallic region of thickness $\ell_{\rm m}$
\begin{align*}
{\bf P}_{\rm pl} ({\bf R})&= \sqrt{\frac{\hbar \epsilon_{0}\omega_{\rm pl}}{2 S \ell_{\rm m}}} \sum_{{\bf Q},\alpha} \left(\Pi_{{\bf -Q}\alpha} + \Pi^{\dagger}_{{\bf Q}\alpha}\right) {\bf u}_{\alpha} e^{-\mi {\bf Q}\cdot {\bf R}} \Theta (-z),
\end{align*}
$H_{\rm mat-pt}$ is derived as
\begin{align*}
H_{\rm mat-pt}= \sum_{\bf q} \Big[a_{\bf q} + a^{\dagger}_{\bf -q} \Big] \Big[\Omega^{\rm sc-pt}_{q} \left(b_{\bf -q} + b^{\dagger}_{\bf q} \right) +  \Omega^{\rm ps-pt}_{q} \left(p_{\bf -q} + p^{\dagger}_{\bf q} \right) + \Omega^{\rm pl-pt}_{q} \left(\pi_{\bf -q} + \pi^{\dagger}_{\bf q} \right)\Big], 
\end{align*}
with the coupling strengths  
\begin{align*}
&\Omega^{\rm pl-pt}_{q} = \sqrt{\frac{c \omega_{\rm pl} \left(q^{2}+\gamma^{2}_{\rm m} \right)}{\gamma_{\rm m}}}, \; \Omega^{\rm sc-pt}_{q}= \nu_{\rm sc} \sqrt{\frac{c f \left(q^{2}+\gamma^{2}_{\rm d} \right)\left(1-e^{-2\gamma_{\rm d}\ell_{\rm d}} \right)}{\epsilon \gamma_{\rm d}\widetilde{\omega}_{\rm sc}}}, \;
\Omega^{\rm ps-pt}_{q}= \nu_{\rm ps} \sqrt{\frac{c (1-f) \left(q^{2}+\gamma^{2}_{\rm d} \right)\left(1-e^{-2\gamma_{\rm d}\ell_{\rm d}} \right)}{\epsilon \gamma_{\rm d}\widetilde{\omega}_{\rm ps}}}.&
\end{align*}
Due to the absence of translational invariance in the $z$ direction, $b_{\bf q}\equiv \sum_{q_{z},\alpha} f_{\alpha} ({\bf Q}) \widetilde{S}_{{\bf Q}\alpha}$, $p_{\bf q}\equiv \sum_{q_{z},\alpha} f_{\alpha} ({\bf Q}) \widetilde{P}_{{\bf Q}\alpha}$, and $\pi_{\bf q}\equiv \sum_{q_{z},\alpha} g_{\alpha} ({\bf Q}) \Pi_{{\bf Q}\alpha}$ are quasi-2d ``bright'' modes, linear superpositions of the 3d modes with different $q_{z}$. The functions $f_{\alpha} ({\bf Q})$ and $g_{\alpha} ({\bf Q})$ are obtained by calculating the overlap integrals between the displacement and the polarization fields in the $z$ direction and read 
\begin{align}
f_{\alpha} ({\bf Q}) = \sqrt{\frac{2 \gamma_{\rm d}}{\ell_{\rm d} \left(q^{2}+ \gamma^{2}_{\rm d}\right) \left(1-e^{-2\gamma_{\rm d}\ell_{\rm d}} \right)}} \times \begin{cases} \frac{-\mi q}{\gamma_{\rm d}-\mi q_{z}} \left(1- e^{\mi q_{z} \ell_{\rm d}}e^{-\gamma_{\rm d} \ell_{\rm d}} \right) \quad \textrm{for} \quad \alpha=z \\ \\
\frac{\gamma_{\rm d}}{\gamma_{\rm d}-\mi q_{z}} \left(1- e^{\mi q_{z} \ell_{\rm d}}e^{-\gamma_{\rm d} \ell_{\rm d}} \right) \quad \textrm{for} \quad \alpha=\parallelsum,
\end{cases}
\label{bright_dd}
\end{align}
and 
\begin{align*}
g_{\alpha} ({\bf Q}) = \sqrt{\frac{2 \gamma_{\rm m}}{\ell_{\rm m} \left(q^{2}+ \gamma^{2}_{\rm m}\right)}} \times \begin{cases} \frac{-\mi q}{\gamma_{\rm m}+\mi q_{z}} \quad \textrm{for} \quad \alpha=z \\ \\
-\frac{\gamma_{\rm m}}{\gamma_{\rm m}+\mi q_{z}} \quad \textrm{for} \quad \alpha=\parallelsum.
\end{cases}
\end{align*}
The prefactors in the previous equations stem from the orthogonality condition $[b_{\bf q},b^{\dagger}_{\bf q'}]=[p_{\bf q},p^{\dagger}_{\bf q'}]=[\pi_{\bf q},\pi^{\dagger}_{\bf q'}]=\delta_{{\bf q},{\bf q'}}$. 

\subsection{Polariton Hamiltonian}
\label{popo_H}

Since light only interacts with quasi-2d bright-modes, we project the matter Hamiltonian Eq.~(\ref{mat_tot}) onto these bright modes and write the polariton Hamiltonian as $H_{\rm pol}=\sum_{\bf q} H^{({\bf q})}_{\rm pol}$ with
\begin{align}
H^{({\bf q})}_{\rm pol} &=\omega_{q} a^{\dagger}_{\bf q} a_{\bf q} + \omega_{\rm pl} \pi^{\dagger}_{\bf q}\pi_{\bf q} +  \Omega^{\rm pl-pt}_{q} \left(\pi_{\bf -q} + \pi^{\dagger}_{\bf q} \right)\left(a_{\bf q} + a^{\dagger}_{\bf -q} \right) +\widetilde{\omega}_{\rm sc} b^{\dagger}_{\bf q}b_{\bf q}+\widetilde{\omega}_{\rm ps} p^{\dagger}_{\bf q}p_{\bf q} \nonumber \\
&+ \Lambda^{\rm sc-ps} \left(b_{\bf -q} + b^{\dagger}_{\bf q} \right)\left(p_{\bf q} + p^{\dagger}_{\bf -q} \right) + \Omega^{\rm sc-pt}_{q} \left(b_{\bf -q} + b^{\dagger}_{\bf q} \right)\left(a_{\bf q} + a^{\dagger}_{\bf -q} \right) +  \Omega^{\rm ps-pt}_{q} \left(p_{\bf -q} + p^{\dagger}_{\bf q} \right)\left(a_{\bf q} + a^{\dagger}_{\bf -q} \right).
\label{Hpol_2D11}
\end{align}
It is also convenient to express $H^{({\bf q})}_{\rm pol}$ in terms of the hybrid phonon modes introduced in Sec.~\ref{mat_HH} as
\begin{align}
H^{({\bf q})}_{\rm pol}&=\omega_{q} a^{\dagger}_{\bf q}a_{\bf q}+\omega_{\rm pl} \pi^{\dagger}_{\bf q}\pi_{\bf q}+ \Omega^{\rm pl-pt}_{q} \left(\pi_{\bf -q} + \pi^{\dagger}_{\bf q} \right)\left(a_{\bf q} + a^{\dagger}_{\bf -q} \right) + \sum_{j} \omega_{j} b^{\dagger}_{{\bf q}j} b_{{\bf q}j} + \Omega^{\rm j-pt}_{q} \left(b_{{\bf -q}j} + b^{\dagger}_{{\bf q}j} \right)\left(a_{\bf q} + a^{\dagger}_{\bf -q} \right),
\label{Hpol_2D}
\end{align}
with the coupling strength between photons and the hybrid mode $j$
\begin{align*}
\Omega^{\rm j-pt}_{q}=\left[\nu_{\rm sc} \sqrt{\frac{f}{\omega_{\rm sc}}} \left(X_{j}-\widetilde{X}_{j} \right) + \nu_{\rm ps} \sqrt{\frac{(1-f)}{\omega_{\rm ps}}} \left(Y_{j}-\widetilde{Y}_{j} \right) \right] \sqrt{\frac{c \left(q^{2}+\gamma^{2}_{\rm d} \right)\left(1-e^{-2\gamma_{\rm d}\ell_{\rm d}} \right)}{\epsilon \gamma_{\rm d}}}.
\end{align*}
The bright modes entering Eq.~(\ref{Hpol_2D}) read $b_{{\bf q}j}= \sum_{q_{z},\alpha} f_{\alpha} ({\bf Q}) \Gamma_{{\bf Q}\alpha j}$. If $q_{z}$ assumes $N$ values, $2N-1$ orthogonal dark modes $d_{{\bf q}j n}= \sum_{q_{z},\alpha} h_{\alpha n} ({\bf Q}) \Gamma_{{\bf Q}\alpha j}$ ($n=1,2,\cdots,2N-1$) can be formally introduced in order to form a complete basis with $[d_{{\bf q}j n},d^{\dagger}_{{\bf q'}j' n'}]=\delta_{{\bf q},{\bf q'}} \delta_{j,j'}\delta_{n,n'}$, $[b_{{\bf q}j},b^{\dagger}_{{\bf q'}j'}]=\delta_{{\bf q},{\bf q'}} \delta_{j,j'}$, and $[b_{{\bf q}j},d^{\dagger}_{{\bf q'}j' n}]=[b_{{\bf q}j},d_{{\bf q'}j' n}]=0$. The polariton Hamiltonian Eq.~(\ref{Hpol_2D}) exhibits $4$ positive eigenvalues $w_{q\zeta}$ in each subspace ${\bf q}$, which can be determined using a self-consistent algorithm working as follows~\cite{todorovY,hagen}. Starting with a given polariton frequency, we use the Helmholtz equation $\epsilon_{i} (w_{q\zeta}) w_{q\zeta}^{2}/c^{2}=q^{2}-\gamma^{2}_{i}$ to determine the field penetration depths entering $H_{\rm pol}$. The dielectric functions $\epsilon_{i} (\omega)$ in the dielectric ($i={\rm d}$) and the metal ($i={\rm m}$) read $\epsilon_{\rm d}=\epsilon$ and $\epsilon_{\rm m} (\omega)=1- \omega^{2}_{\rm pl}/\omega^{2}$. The polariton Hamiltonian is then diagonalized numerically and the algorithm is repeated with the lowest eigenvalue as a new polariton frequency until convergence is reached. 

\begin{figure}[ht]
\centerline{\includegraphics[width=0.95\columnwidth]{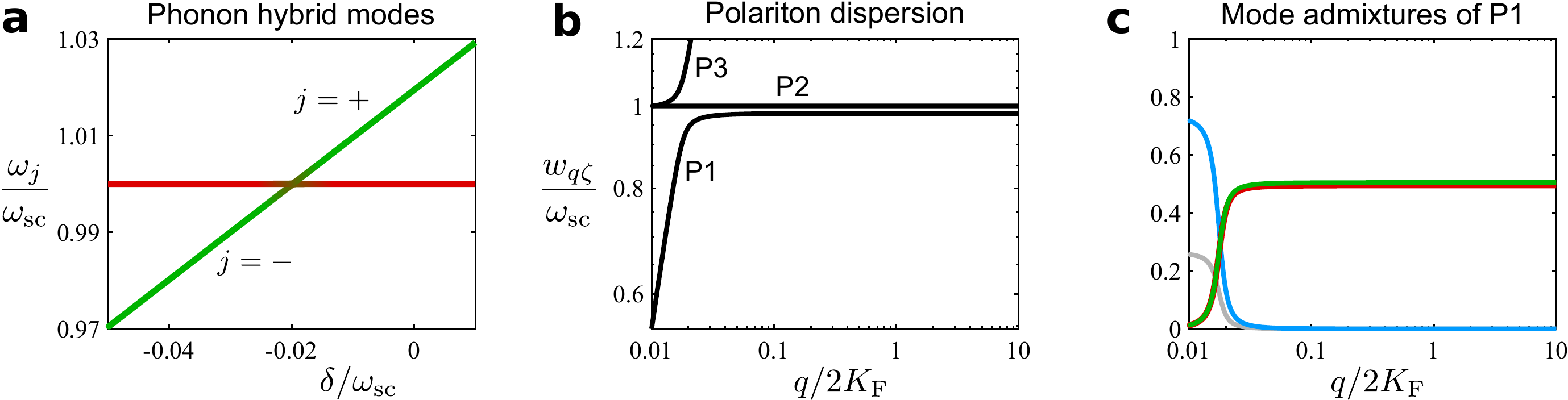}}
\caption{\textbf{a}, Normalized frequencies of the phonon hybrid modes $\omega_{j}$ given by Eq.~(\ref{E_hyb}) as a function of the detuning $\delta =\omega_{\rm ps}-\omega_{\rm sc}$, together with the mode admixtures (green for PS and red for SC). \textbf{b}, Frequency dispersion $w_{q \zeta}$ of the $3$ lowest polaritons $\zeta={\rm P1},{\rm P2},{\rm P3}$ as a function of the in-plane wave vector $q$ (normalized by the Fermi wave vector $K_{\rm F}$) for $\delta\approx -0.02\omega_{\rm sc}$. \textbf{c}, Modes admixtures of the lowest polariton P1 versus $q/2K_{\rm F}$ for $\delta\approx -0.02\omega_{\rm sc}$. Blue, grey, green, and red correspond to photons, plasmons, PS phonons, and SC phonons, respectively. Parameters are $f=0.02$, $\nu_{\rm sc}=0.01 \omega_{\rm sc}$, $\nu_{\rm ps}=0.2 \omega_{\rm sc}$, $K_{\rm F}=0.1 {\rm nm}^{-1}$, $\ell_{\rm d}=20 {\rm nm}$, $\hbar \omega_{\rm sc}=0.18 {\rm eV}$, and $\omega_{\rm pl}=9 {\rm eV}$.} 
\label{fig_hopfield}
\end{figure}

The frequency dispersion of the three lowest polaritons is represented in Fig.~\ref{fig_hopfield}\textbf{b} together with the mode admixtures of the lowest branch P1 (see Fig.~\ref{fig_hopfield}\textbf{c}) as a function of the in-plane wave vector $q$. When the two shifted phonons are in resonance ($\widetilde{\omega}_{\rm sc}=\widetilde{\omega}_{\rm ps}$) which occurs for a detuning $\delta=\omega_{\rm ps}-\omega_{\rm sc}\approx -0.02\omega_{\rm sc}$, the polariton P1 is composed of $50\%$ PS phonons and $50\%$ SC phonons at large wave vectors $q \sim K_{\rm F}$. Note that the ``polaritonic'' gap between P1 and P2 originates from the compensation between the depolarization shift of the phonon modes and the dispersive shift induced by light-matter interactions at large $q$. This can be shown by eliminating the photon operators $a_{\bf q}$ and $a^{\dagger}_{\bf q}$ in Eq.~(\ref{Hpol_2D11}) using second-order perturbation theory. Note that the ``counter-rotating'' terms $\propto b_{\bf -q} a_{\bf q}, \propto p_{\bf -q} a_{\bf q}$ and the hermitian conjugates are crucial to the existence of the polaritonic gap.

\subsection{Enhancement of the electron-phonon coupling}
\label{elec_pho}

The coupling between electrons in the $t_{1u}$ band of ${\rm Rb}_{3} {\rm C}_{60}$ and the on-ball phonon mode of C$_{60}$ is described by the Hamiltonian
\begin{align}
H_{\rm el-pn}= \sum_{\bf K} \xi_{K} c^{\dagger}_{\bf K} c_{\bf K} + V \sum_{{\bf K},{\bf Q},\alpha} c^{\dagger}_{\bf K} c_{{\bf K}-{\bf Q}} \left(S_{{\bf Q}\alpha}+ S^{\dagger}_{{\bf -Q}\alpha}\right),  
\label{elec_phfinal}
\end{align}
where the electron-phonon matrix element $V$ does not depend on the wave vector ${\bf Q}$ at the lowest order, as shown in Ref.~[\onlinecite{PhysRevB.44.12106}]. The fermionic operator $c_{\bf K}$ ($c^{\dagger}_{\bf K}$) annihilates (creates) a spin-less electron with 3d wave vector ${\bf K}$ and energy $\xi_{K}$ ($K = \vert {\bf K} \vert$) relative to the Fermi energy. For simplicity, we assume a spherical Fermi surface providing $\xi_{K} =\hbar^{2} (K^{2}- K^{2}_{\rm F})/(2 m)$, with $m$ the electron effective mass. 

Electron-phonon interactions are characterised by the dimensionless coupling parameter $\lambda$, which quantifies the electron mass renormalization due to the coupling to phonons. At zero temperature, this parameter is defined as
\begin{align}
\lambda= \frac{1}{N(0)}\sum_{\bf K} \delta (\xi_{K}) \Re \left(-\partial_{\omega} \overline{\Sigma}_{\bf K} (\omega) \vert_{\omega=0} \right),
\label{lambda_eq}
\end{align}
where $\Re$ stands for real part, $N(0)=\sum_{\bf K} \delta (\xi_{K})=\frac{S \ell_{\rm d} m K_{\rm F}}{2 \pi^2 \hbar}$ is the electron density of states at the Fermi level, and $\partial_{\omega} \overline{\Sigma}_{\bf K} (\omega) \vert_{\omega=0}$ denotes the frequency derivative of the retarded electron self-energy $\overline{\Sigma}_{\bf K} (\omega)$ evaluated at $\omega=0$. The latter is derived from the Hamiltonian Eq.~(\ref{elec_phfinal}) as~\cite{mahan}  
\begin{align}
\Sigma_{\bf K}(\omega)= \mi V^{2} \sum_{{\bf Q},\alpha} \int \!\! \frac{d\omega'}{2\pi} \mathcal{G}_{{\bf K}-{\bf Q}} (\omega+\omega') \mathfrak{S}_{Q\alpha} (\omega'), 
\label{self_eq0}
\end{align}
where $\mathcal{G}_{\bf K} (\tau)=-\mi \int \! d\tau e^{\mi\omega \tau} \langle c_{\bf K} (\tau) c^{\dagger}_{\bf K} (0) \rangle$ denotes the electron Green's function (GF), and $\mathfrak{S}_{Q\alpha} (\omega)=-\mi \int \! d\tau e^{\mi\omega \tau} \langle \mathcal{S}_{{\bf Q}\alpha} (\tau)\mathcal{S}_{{\bf -Q}\alpha} (0) \rangle$ the phonon GF with $\mathcal{S}_{{\bf Q}\alpha}=S_{{\bf Q}\alpha}+S^{\dagger}_{{\bf -Q}\alpha}$. 

In the absence of phonon-photon coupling ($\nu_{\rm sc}=\nu_{\rm ps}=0$), SC phonons enter the Hamiltonian $H_{\rm pol}$ only via the free contribution $\omega_{\rm sc} \sum_{{\bf Q},\alpha} S^{\dagger}_{{\bf Q}\alpha}S_{{\bf Q}\alpha}$, and $\mathfrak{S}_{Q\alpha}(\omega)=2\omega_{\rm sc}/(\omega^{2}-\omega^{2}_{\rm sc})$ thus corresponds to the non-interacting SC phonon GF. Using this expression and approximating $\mathcal{G}_{\bf K}$ by the non-interacting electron GF $\mathcal{G}^{0}_{\bf K} (\omega)=1/(\omega -\xi_{K})$ in Eq.~(\ref{self_eq0}), one can show that Eq.~(\ref{lambda_eq}) provides~\cite{mahan} 
\begin{align*}
\lambda_{0}= \frac{2 V^{2}}{N(0)\omega_{\rm sc}} \sum_{\bf K} \delta (\xi_{K}) \sum_{{\bf Q},\alpha} \delta (\xi_{{\bf K}-{\bf Q}}) =\frac{4 V^{2} N(0)}{\omega_{\rm sc}}.
\end{align*}

For $\nu_{\rm sc}\neq 0$ and $\nu_{\rm ps}\neq 0$, the phonon dynamics is governed by the Hamiltonian $H_{\rm pol}$, which includes the coupling of SC phonons to photons and PS phonons. The projection of the electron-phonon coupling Hamiltonian Eq.~(\ref{elec_phfinal}) onto the bright and dark hybrid phonon modes defined in Sec.~\ref{popo_H} allows to decompose the self-energy as $\Sigma_{\bf K} (\omega)=\Sigma^{\rm (B)}_{\bf K} (\omega) +\Sigma^{\rm (D)}_{\bf K} (\omega)$, where
\begin{align}
\Sigma^{\rm (B)}_{\bf K} (\omega)&= \mi V^{2} \sum_{{\bf Q},\alpha} \sum_{j} \left(X_{j}-\widetilde{X}_{j} \right)^{2} \vert f_{\alpha} ({\bf Q}) \vert^{2} \int \!\! \frac{d\omega'}{2\pi} \mathcal{G}_{{\bf K}-{\bf Q}} (\omega+\omega') \mathfrak{B}_{q j} (\omega') 
\label{self_eq236} \\
\Sigma^{\rm (D)}_{\bf K} (\omega)&= \mi V^{2} \sum_{{\bf Q},\alpha} \sum_{j,n} \left(X_{j}-\widetilde{X}_{j} \right)^{2} \vert h_{\alpha n} ({\bf Q}) \vert^{2} \int \!\! \frac{d\omega'}{2\pi} \mathcal{G}_{{\bf K}-{\bf Q}} (\omega+\omega') \mathfrak{D}_{q j n} (\omega').
\label{self_eq237}
\end{align}
Here, $\mathfrak{B}_{q j} (\omega)=-\mi \int \! d\tau e^{\mi \omega \tau} \langle \mathcal{B}_{{\bf q}j} (\tau) \mathcal{B}_{{\bf -q} j} (0) \rangle$ and $\mathfrak{D}_{q j n} (\omega)=-\mi \int \! d\tau e^{\mi \omega \tau} \langle \mathcal{D}_{{\bf q}j n} (\tau) \mathcal{D}_{{\bf -q} j n} (0) \rangle$ denote the bright and dark mode GFs, respectively, with $\mathcal{B}_{{\bf q} j}=b_{{\bf q} j}+b^{\dagger}_{{\bf -q} j}$ and $\mathcal{D}_{{\bf q} j n}=d_{{\bf q} j n}+d^{\dagger}_{{\bf -q} j n}$. 

\vspace{3mm}

We now decompose the electron-phonon coupling parameter into its bright and dark mode contributions as $\lambda=\lambda^{\rm (B)}+\lambda^{\rm (D)}$ for $\nu_{\rm sc}\neq 0$ and $\nu_{\rm ps}\neq 0$, and $\lambda_{0}=\lambda^{\rm (B)}_{0}+\lambda^{\rm (D)}_{0}$ for $\nu_{\rm sc}=\nu_{\rm ps}= 0$. In the latter case, the bright and dark mode GFs simply read $\mathfrak{B}_{q j} (\omega)=\mathfrak{D}_{q j n} (\omega)=2\omega_{j}/(\omega^{2}-\omega^{2}_{j})$ with $\omega_{-}=\omega_{\rm ps}$ and $\omega_{+}=\omega_{\rm sc}$ (assuming $\omega_{\rm ps}< \omega_{\rm sc}$). The hybrid mode admixtures read $X_{-}=\widetilde{X}_{-}=\widetilde{X}_{+}=0$ and $X_{+}= 1/\sqrt{2}$. Note that the $\pm$ signs are reversed for $\omega_{\rm ps}> \omega_{\rm sc}$. Using these GFs together with $\mathcal{G}^{0}_{\bf K} (\omega)=1/(\omega -\xi_{K})$ in Eqs.~(\ref{self_eq236}) and (\ref{self_eq237}) yields
\begin{align}
\lambda^{\rm (B)}_{0}&= \frac{2 V^{2}}{N(0)\omega_{\rm sc}} \sum_{\bf K} \delta (\xi_{K}) \sum_{{\bf Q},\alpha} \vert f_{\alpha} ({\bf Q}) \vert^{2}\delta (\xi_{{\bf K}-{\bf Q}}) \nonumber \\
\lambda^{\rm (D)}_{0}&= \frac{2 V^{2}}{N(0)\omega_{\rm sc}} \sum_{\bf K} \delta (\xi_{K}) \sum_{{\bf Q},\alpha,n} \vert h_{\alpha n} ({\bf Q}) \vert^{2}\delta (\xi_{{\bf K}-{\bf Q}}).
\label{lambda_eq1_BD}
\end{align}
For $\nu_{\rm sc}\neq 0$ and $\nu_{\rm ps}\neq 0$, the dark modes (which do not interact with photons) become hybrid phonon modes. Using again the same procedure with $\mathfrak{D}_{q j n} (\omega)=2\omega_{j}/(\omega^{2}-\omega^{2}_{j})$ in Eq.~(\ref{self_eq237}), we obtain
\begin{align}
\lambda^{\rm (D)}=  \sum_{j} \frac{2 V^{2}}{N(0) \omega_{j}} \left(X_{j}-\widetilde{X}_{j} \right)^{2} \sum_{\bf K} \delta (\xi_{K}) \sum_{{\bf Q},\alpha,n} \vert h_{\alpha n} ({\bf Q}) \vert^{2}\delta (\xi_{{\bf K}-{\bf Q}}).
\label{lambda_eq2_BD}
\end{align}
At zeroth order in $\frac{\nu_{\rm sc}}{\omega_{\rm sc}}\frac{\nu_{\rm ps}}{\omega_{\rm ps}}\sqrt{f(1-f)}$ and $\frac{\nu^{2}_{\rm sc}f}{\omega^{2}_{\rm sc}}$, one finds $X_{-}=X_{+}\approx 1/\sqrt{2}$, $\widetilde{X}_{-}=\widetilde{X}_{+}\approx 0$, and $\omega_{-}\approx \omega_{+}\approx \omega_{\rm sc}$ when the two shifted phonons are in resonance ($\widetilde{\omega}_{\rm sc}=\widetilde{\omega}_{\rm ps}$). These approximations used in Eqs.~(\ref{lambda_eq1_BD}) and (\ref{lambda_eq2_BD}) yield $\lambda^{\rm (D)}\approx \lambda^{\rm (D)}_{0}$. 

We use the equation of motion theory~\cite{hagen} with the Hamiltonian $H_{\rm pol}$ to calculate the bright mode GF $\mathfrak{B}_{q j} (\omega)$ for $\nu_{\rm sc}\neq 0$ and $\nu_{\rm ps}\neq 0$. The latter is provided by the solution of the Dyson equation 
\begin{align}
\mathfrak{B}_{q j} (\omega)=\mathfrak{B}^{0}_{j} (\omega) + \sum_{j'} \mathfrak{B}^{0}_{j} (\omega) \Phi^{jj'}_{q} (\omega) \mathfrak{B}_{q j'} (\omega),
\label{pho_GFGf}
\end{align}
where the bright mode self-energy reads 
\begin{align*}
\Phi^{jj'}_{q} (\omega)=\frac{\omega^{2}_{q} \left(\omega^{2}- \omega^{2}_{\rm pl} \right) \Omega^{\rm j-pt}_{q} \Omega^{\rm j'-pt}_{q}}{2 c \alpha_{q} \left(\omega^{2} - \omega^{2}_{q} \right) \left(\omega^{2}- \omega^{2}_{\rm pl} \right) - 2 \omega^{2}_{q} \omega_{\rm pl} \left(\Omega^{\rm pl-pt}_{q} \right)^{2}}.
\end{align*}
Solving Eq.~(\ref{pho_GFGf}), the bright mode GF can be put in the form 
\begin{align}
\mathfrak{B}_{q j} (\omega)=\sum_{j'} (1- \delta_{j,j'}) \frac{2\omega_{j} R^{jj'}_{q} (\omega)}{c \alpha_{q} \prod_{\zeta} \left(\omega^{2} - w^{2}_{q\zeta} \right)},
\label{pho_GFGf_final}
\end{align}
where $w_{q\zeta}$ are the polariton frequencies, eigenvalues of the Hamiltonian Eq.~(\ref{Hpol_2D}) and 
\begin{align*}
R^{jj'}_{q} (\omega)= \Big[\omega^{2} - \omega^{2}_{j'} \Big] \Big[c\alpha_{q}\left(\omega^{2} - \omega^{2}_{q} \right) \left(\omega^{2}- \omega^{2}_{\rm pl} \right) - \omega^{2}_{q} \omega_{\rm pl} \left(\Omega^{\rm pl-pt}_{q} \right)^{2} \Big] + \omega_{j'} \omega^{2}_{q} \left(\omega^{2}- \omega^{2}_{\rm pl} \right) \Omega^{\rm j'-pt}_{q} \left(\Omega^{\rm j-pt}_{q} - \Omega^{\rm j'-pt}_{q}\right).  
\end{align*}
Using the GF Eq.~(\ref{pho_GFGf_final}) and $\mathcal{G}^{0}_{\bf K} (\omega)=1/(\omega -\xi_{K})$, we calculate the self-energy Eq.~(\ref{self_eq236}), which together with Eqs.~(\ref{lambda_eq}) and (\ref{lambda_eq1_BD}) leads to
\begin{align}
\frac{\Delta\lambda}{\lambda_{0}}\equiv \frac{\lambda-\lambda_{0}}{\lambda_{0}}= \frac{1}{2N^{2}(0)} \sum_{\bf K} \delta (\xi_{K}) \sum_{{\bf Q},\alpha} \left(\varphi_{q} - 1 \right) \vert f_{\alpha} ({\bf Q}) \vert^{2} \delta (\xi_{{\bf K}-{\bf Q}}).
\label{rel}
\end{align}
The function
\begin{align}
\varphi_{q} = \sum_{\zeta} \sum_{j,j'}\left(1-\delta_{j,j'} \right)\frac{\omega_{j}\omega_{\rm sc} R^{j j'}_{q} (w_{q\zeta}) \left(X_{j}-\widetilde{X}_{j} \right)^{2}}{w^{2}_{q\zeta} c \alpha_{q}  \prod_{\zeta'\neq \zeta} \left(w^{2}_{q\zeta} - w^{2}_{q\zeta'} \right)}
\label{sum_phi}
\end{align} 
describes the renormalization of the SC phonon energy due to the coupling to PS phonons and photons. The enhancement of the electron-phonon coupling $\frac{\Delta\lambda}{\lambda_{0}}$ is computed by transforming the summation over ${\bf Q}$ in Eq.~(\ref{rel}) into an integral and introducing the new variable $x=Q/(2 K_{\rm F})$. This yields 
\begin{align*}
\frac{\Delta\lambda}{\lambda_{0}}= \int^{1}_{0} \! dx \,x F(x),
\end{align*}
where $F (x)$ is a dimensionless function defined as $F (\frac{q}{2K_{\rm F}},\frac{q_{z}}{2K_{\rm F}})= \sum_{\alpha} \left(\varphi_{q} -1 \right) \vert f_{\alpha} (Q) \vert^{2}$ with the replacements $q/2K_{\rm F} \to x\sqrt{1-x^{2}}$ and $q_{z}/2K_{\rm F}\to x^{2}$. We find that the contribution of the lowest polariton $\zeta={\rm P1}$ largely dominates the sum in Eq.~(\ref{sum_phi}), and since $w_{q {\rm P1}}$ is constant for $q\sim K_{\rm F} \gg \omega_{\rm sc}/c$, the function $\varphi_{q}$ can also be considered as constant for large wave vectors $q\sim K_{\rm F}$. In this case, one can use Eq.~(\ref{bright_dd}) with $\gamma_{\rm d}\sim q$ (for $q\sim K_{\rm F} \gg \omega_{\rm pl}/c$) to put the relative enhancement of $\lambda$ in the simple form 
$$\frac{\Delta\lambda}{\lambda_{0}}\approx \frac{\pi (\varphi-1)}{4 K_{\rm F} \ell_{\rm d}}$$ in a 3d configuration for which $K_{\rm F}\ell_{\rm d} \gg 1$.

\subsection{Fitting of the ATR measurements}
\label{fit_sys}

\begin{figure}[ht]
\centerline{\includegraphics[width=1\columnwidth]{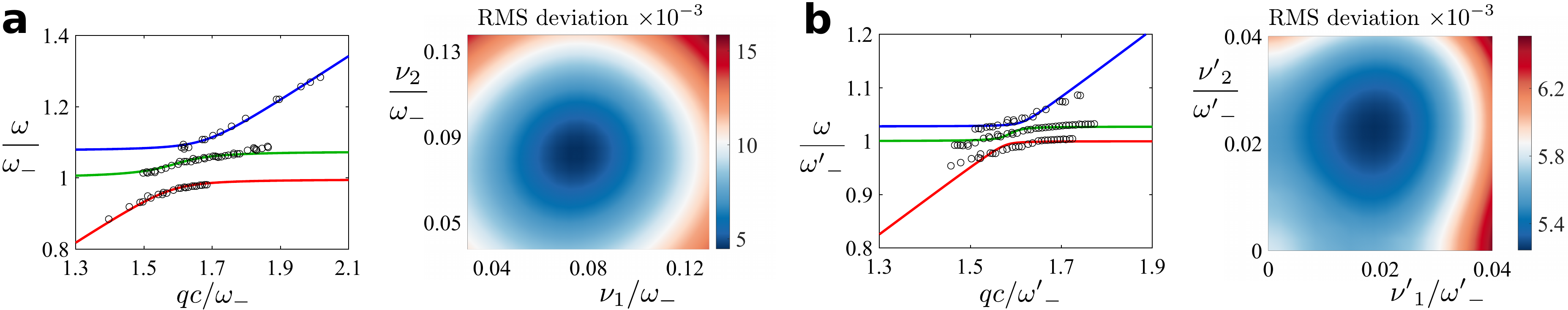}}
\caption{Normalized energy dispersion of the surface polaritons (PS film on Au) obtained with the theoretical model (colored lines) in the low-frequency (\textbf{a}) and high-frequency (\textbf{b}) ranges, for $\nu_{1}=0.074\omega_{-}$, $\nu_{2}=0.082\omega_{-}$, ${\nu'}_{1}=0.019{\omega'}_{-}$, and ${\nu'}_{2}=0.024{\omega'}_{-}$ (best fit). Black circles are experimental points. The right panels in \textbf{a} and \textbf{b} show the RMS deviations as a function of the ionic plasma frequencies. The best fits correspond to the minima of the RMS deviation.} 
\label{fig_fit}
\end{figure}

We now use our model to extract the typical ionic plasma frequencies of PS from the ATR spectra of the PS film on Au. The transmission spectrum of PS features $4$ infrared-active modes in the spectral range of interest, i.e. close to the phonon modes involved in superconductivity for ${\rm Rb}_{3} {\rm C}_{60}$ and ${\rm YBCO}$. We first focus on the two modes with frequencies $\omega_{1}$ and $\omega_{2}$ located in the lower part of the spectrum $\omega \sim 700 {\rm cm}^{-1}$, and introduce the polarization fields 
\begin{align*}
{\bf P}_{n} ({\bf R})&= \nu_{\rm n}\sqrt{\frac{\hbar \epsilon_{0}\epsilon}{2\omega_{\rm n} S \ell_{\rm d}}} \sum_{{\bf Q},\alpha} \left(P_{{\bf -Q}\alpha n} + P^{\dagger}_{{\bf Q}\alpha n}\right) {\bf u}_{\alpha} e^{-\mi {\bf Q}\cdot {\bf R}} \Theta (z)
\end{align*}
with $n=1,2$. We proceed as in Secs.~\ref{mat_HH}, \ref{pho_H}, and \ref{light_ma}, and derive the polariton Hamiltonian including both the light-matter and dipole-dipole couplings. The latter is responsible for hybridization between the two vibrational modes, which leads to two new modes with frequencies $\omega_{\pm}$. Since the transmission spectrum shown in Fig.~2\textbf{a} of the main text involves optical wave vectors $q \lesssim \omega_{n}/c$, one can associate the transmittance minima to these hybrid modes (top of the polaritonic gap), i.e. $\omega_{-}=700 {\rm cm}^{-1}$ and $\omega_{+}=755 {\rm cm}^{-1}$. We diagonalize the polariton Hamiltonian and compare the resulting eigenvalues $\omega^{\rm th}_{q_{i},\zeta}$ with the experimental points $\omega^{\rm exp}_{q_{i},\zeta}$ by computing the root mean square (RMS) deviation
\begin{align*}
{\rm RMS} =\sqrt{\frac{\sum_{i=1}^{N} \sum_{\zeta=1,2,3} \left(\omega^{\rm exp}_{q_{i},\zeta} - \omega^{\rm th}_{q_{i},\zeta} \right)^{2}}{2 N}},
\end{align*}  
where $i=1,\cdots,N$ runs over all experimental points, and $\zeta$ denotes the three lowest polaritons. The adjustable parameters of the fit are the ionic plasma frequencies $\nu_{1}$ and $\nu_{2}$ associated to the vibrational modes $n=1,2$. The same procedure is applied for the two other modes in the upper part of the spectrum (see Fig.~3 \textbf{a} of the main text) at ${\omega'}_{-}=1450 {\rm cm}^{-1}$, and ${\omega'}_{+}=1490 {\rm cm}^{-1}$, with the ionic plasma frequencies $\nu'_{1}$ and $\nu'_{2}$. The energy dispersion in both parts of the spectrum are represented in Fig.~\ref{fig_fit} together with the RMS deviation. The latter is minimum for $\nu_{1}=0.074 \omega_{-}$, $\nu_{2}=0.082\omega_{-}$, $\nu'_{1}=0.019{\omega'}_{-}$, and $\nu'_{2}=0.024{\omega'}_{-}$. 

In metallic crystals (including superconductors), the ionic plasma frequency associated to infrared-active phonon modes is typically small due to the screening of the effective ionic charge by free electrons~\cite{PhysRevLett.103.116804}. In ${\rm K}_{3} {\rm C}_{60}$, one can estimate the ionic plasma frequency associated to the $t_{1u}(4)$ phonon~\cite{Mitrano2016} (on-ball mode of ${\rm C}_{60}$) to be $\nu \sim 3 {\rm THz}$, which provides $\nu \sim 0.01\omega_{-}$ typically smaller than the values for PS found above. This justifies our assumption that $\nu_{\rm ps} > \nu_{\rm sc}$.
  
\bibliography{supra_light_supp}